\PassOptionsToPackage{unicode}{hyperref}
\PassOptionsToPackage{hyphens}{url}
\documentclass[
]{article}
\usepackage{amsmath,amssymb}
\usepackage{lmodern}
\usepackage{iftex}
\ifPDFTeX
  \usepackage[T1]{fontenc}
  \usepackage[utf8]{inputenc}
  \usepackage{textcomp} 
\else 
  \usepackage{unicode-math}
  \defaultfontfeatures{Scale=MatchLowercase}
  \defaultfontfeatures[\rmfamily]{Ligatures=TeX,Scale=1}
\fi
\IfFileExists{upquote.sty}{\usepackage{upquote}}{}
\IfFileExists{microtype.sty}{
  \usepackage[]{microtype}
  \UseMicrotypeSet[protrusion]{basicmath} 
}{}
\makeatletter
\@ifundefined{KOMAClassName}{
  \IfFileExists{parskip.sty}{%
    \usepackage{parskip}
  }{
    \setlength{\parindent}{0pt}
    \setlength{\parskip}{6pt plus 2pt minus 1pt}}
}{
  \KOMAoptions{parskip=half}}
\makeatother
\usepackage{xcolor}
\usepackage{longtable,booktabs,array}
\usepackage{multirow}
\usepackage{calc} 
\usepackage{etoolbox}
\makeatletter
\patchcmd\longtable{\par}{\if@noskipsec\mbox{}\fi\par}{}{}
\makeatother
\IfFileExists{footnotehyper.sty}{\usepackage{footnotehyper}}{\usepackage{footnote}}
\makesavenoteenv{longtable}
\usepackage{graphicx}
\makeatletter
\def\maxwidth{\ifdim\Gin@nat@width>\linewidth\linewidth\else\Gin@nat@width\fi}
\def\maxheight{\ifdim\Gin@nat@height>\textheight\textheight\else\Gin@nat@height\fi}
\makeatother
\setkeys{Gin}{width=\maxwidth,height=\maxheight,keepaspectratio}
\makeatletter
\def\fps@figure{htbp}
\makeatother
\ifLuaTeX
  \usepackage{selnolig}  
\fi
\IfFileExists{bookmark.sty}{\usepackage{bookmark}}{\usepackage{hyperref}}
\IfFileExists{xurl.sty}{\usepackage{xurl}}{} 
\hypersetup{
  hidelinks,
  pdfcreator={LaTeX via pandoc}}

\author{}
\date{}

\begin{document}

Title:

\textbf{Artificial Intelligence in Fetal Resting-State Functional MRI
Brain Segmentation: A Comparative Analysis of 3D UNet, VNet, and
HighRes-Net Models}

Farzan Vahedifard *\footnote{Department of Diagnostic Radiology and
  Nuclear Medicine, Rush Medical College. Email:
  Farzan\_vahedifard@Rush.edu}; Xuchu Liu \footnote{Department of
  Diagnostic Radiology and Nuclear Medicine, Rush Medical College.
  Email: xuchu\_liu@rush.edu} , Mehmet Kocak\footnote{Associate
  Professor, Department of Diagnostic Radiology and Nuclear Medicine,
  Rush Medical College. Email: mehmet\_kocak@rush.edu} , H. Asher
Ai\footnote{Division for Diagnostic Medical Physics, Department of
  Radiology and Nuclear Medicine at Rush University Medical Center.
  Email: hua\_a\_ai@rush.edu} , Mark Supanich \footnote{Division for
  Diagnostic Medical Physics, Department of Radiology and Nuclear
  Medicine at Rush University Medical Center, Email:
  mark\_supanich@rush.edu}; Christopher Sica. \footnote{Technical \&
  Operations Director MRI, Rush Imaging Research Core, Rush Medical
  College , Email: christopher\_sica@rush.edu}, Kranthi K Marathu
\footnote{Department of Diagnostic Radiology and Nuclear Medicine, Rush
  Medical College. Email: kranthi\_k\_marathu@rush.edu}; Seth Adler
\footnote{Department of Diagnostic Radiology and Nuclear Medicine, Rush
  Medical College. Email: seth\_adler@rush.edu},
Maysam~Orouskhani~\footnote{Department of Radiology, University of
  Washington, Seattle, WA, USA. Email: maysam@uw.edu}; Sharon Byrd
\footnote{Professor and Chairperson, Department of Diagnostic Radiology
  and Nuclear Medicine, Rush Medical College. Email:
  sharon\_byrd@rush.edu}

The funding of this project is from the Colonel Robert McCormick
Diagnostic Chair Spending fund No. 840152-03 of Rush University Medical
Center in Chicago, Illinois.

\textbf{Artificial Intelligence in Fetal Resting-State Functional MRI
Brain Segmentation: A Comparative Analysis of 3D UNet, VNet, and
HighRes-Net Models}

\textbf{Abstract:}

\textbf{Introduction:} Fetal resting-state functional magnetic resonance
imaging (rs-fMRI) is a rapidly evolving field that provides valuable
insight into brain development before birth. Accurate segmentation of
the fetal brain from the surrounding tissue in nonstationary 3D brain
volumes poses a significant challenge in this domain. Current available
tools have 0.15 accuracy.

\textbf{Aim:} This study introduced a novel application of artificial
intelligence (AI) for automated brain segmentation in fetal brain fMRI,
magnetic resonance imaging (fMRI). Open datasets were employed to train
AI models, assess their performance, and analyze their capabilities and
limitations in addressing the specific challenges associated with fetal
brain fMRI segmentation.

\textbf{Method:} We utilized an open-source fetal functional MRI (fMRI)
dataset consisting of 160 cases (reference: fetal-fMRI - OpenNeuro). An
AI model for fMRI segmentation was developed using a 5-fold
cross-validation methodology. Three AI models were employed: 3D UNet,
VNet, and HighResNet. Optuna, an automated hyperparameter-tuning tool,
was used to optimize these models.

\textbf{Results and Discussion:} The Dice scores of the three AI models
(VNet, UNet, and HighRes-net) were compared, including a comparison
between manually tuned and automatically tuned models using Optuna. Our
findings shed light on the performance of different AI models for fetal
resting-state fMRI brain segmentation. Although the VNet model showed
promise in this application, further investigation is required to fully
explore the potential and limitations of each model, including the
HighRes-net model. This study serves as a foundation for further
extensive research into the applications of AI in fetal brain fMRI
segmentation.


\hypertarget{introduction}{%
\section{Introduction:}\label{introduction}}

Fetal resting-state functional magnetic resonance imaging (rs-fMRI) is a
flourishing domain in neuroscience and developmental biology that offers
a unique window into the intricate processes that shape brain
development before birth (1, 2). This non-invasive imaging technique has
emerged as a powerful tool for studying the formation and evolution of
the brain\textquotesingle s network during this critical phase,
providing real-time insights into neural activity and connectivity (3,
4). The application of rs-fMRI in the assessment of brain functionality
during fetal development is important because it allows for the
evaluation of actual brain responsiveness and functioning, going beyond
the structural brain information provided by traditional imaging
techniques (4).

However, fetal fMRI presents unique challenges that limit its use in
clinical and research settings (5). Issues related to movement and
technical scanning, along with limitations in providing reliable and
spatially accurate details of fetal brain activity, have hindered the
effectiveness of fMRI in this context (5). Furthermore, the dynamic
nature of the fetal brain, susceptibility artifacts introduced by
surrounding maternal tissues, and physiological noise from both the
mother and fetus add complexity to the image processing and analysis
procedure (5, 6). Additionally, the lack of standardization in the
orientation and shape of the fetal brain poses challenges for accurate
segmentation (7).

To address these challenges and improve the reliability and efficiency
of fetal brain segmentation in rs-fMRI, Artificial Intelligence (AI)
techniques have gained prominence (8). AI models, particularly
Convolutional Neural Networks (CNNs), have succeeded in image analysis
tasks including brain segmentation (8-12). However, most existing AI
models for brain segmentation have been developed for adult brains and
do not perform well when applied to fetal brain data (8). Therefore, a
comparative analysis of AI models specifically tailored for fetal
rs-fMRI brain segmentation is necessary to determine their effectiveness
and to identify the most suitable approach.

This study aims to compare the performance of three AI models in fetal
resting-state fMRI brain segmentation: 3D UNet, VNet, and HighRes-Net
(8). By utilizing open datasets and conducting a comprehensive
evaluation of these models, this study aims to assess their accuracy,
efficiency, and robustness in dealing with the unique challenges of
fetal brain segmentation (8). These findings will contribute to the
development of improved AI-based segmentation methods for fetal rs-fMRI,
enhancing our understanding of fetal brain development, and facilitating
clinical interventions for neurodevelopmental disorders (8).

In subsequent sections, this paper will provide a review of the existing
literature on fetal fMRI and the challenges associated with brain
segmentation in this context, followed by a detailed methodology
description, an analysis of the results, and a discussion of the
implications of the study\textquotesingle s findings (1, 2, 7, 8). By
examining the potential of AI-based segmentation models in fetal fMRI,
this research aims to pave the way for advancements in the field and
improve our understanding of fetal brain development and maturation (8).

\textbf{Aim:} This study introduces a novel application of artificial
intelligence (AI) for automated brain segmentation in fetal brain
functional MRI (fMRI). Open datasets were employed to train AI models,
assess their performance, and analyze their capabilities and limitations
in addressing the specific challenges associated with fetal brain fMRI
segmentation. Each model has unique architectural and functional
characteristics, which make it suitable for specific applications. For
instance, the VNet model demonstrated superior accuracy in our study,
whereas the 3D UNet model showed a more consistent performance.
Additionally, despite its lower scores, the HighRes-net model might hold
potential advantages in other aspects that were not investigated in this
study.

The limitations of existing models underscore the need for further
research to explore the full potential and weaknesses of each model,
especially for application to fetal functional MRI data. Our work forms
the foundation for more extensive research into the applications of AI
in fetal brain fMRI segmentation, promising new insights into brain
development before birth.

\hypertarget{method}{%
\section{Method}\label{method}}

In this study, we introduced a novel application of AI for automated
brain segmentation in fetal brain fMRI. We utilized an open-source fetal
functional MRI (fMRI) dataset comprising 160 cases from OpenNeuro. Our
approach involved the development of an artificial intelligence (AI)
model for fMRI segmentation (Figure 1) using a 5-fold cross-validation
approach. Table 1 shows the distribution of data.

Left and middle columns: Original dataset distribution (Details of how
data were separated into training, validation, and test sets. Two
iterations of the auto mask CNN model were run to compare single-site
results (iteration 1) with multi-site results (iteration 2). In our
model we didn't use the Yale test set, as we didn't have access to
that.\\
Right Column: Our cross validation.

\includegraphics[width=6.64163in,height=2.40442in]{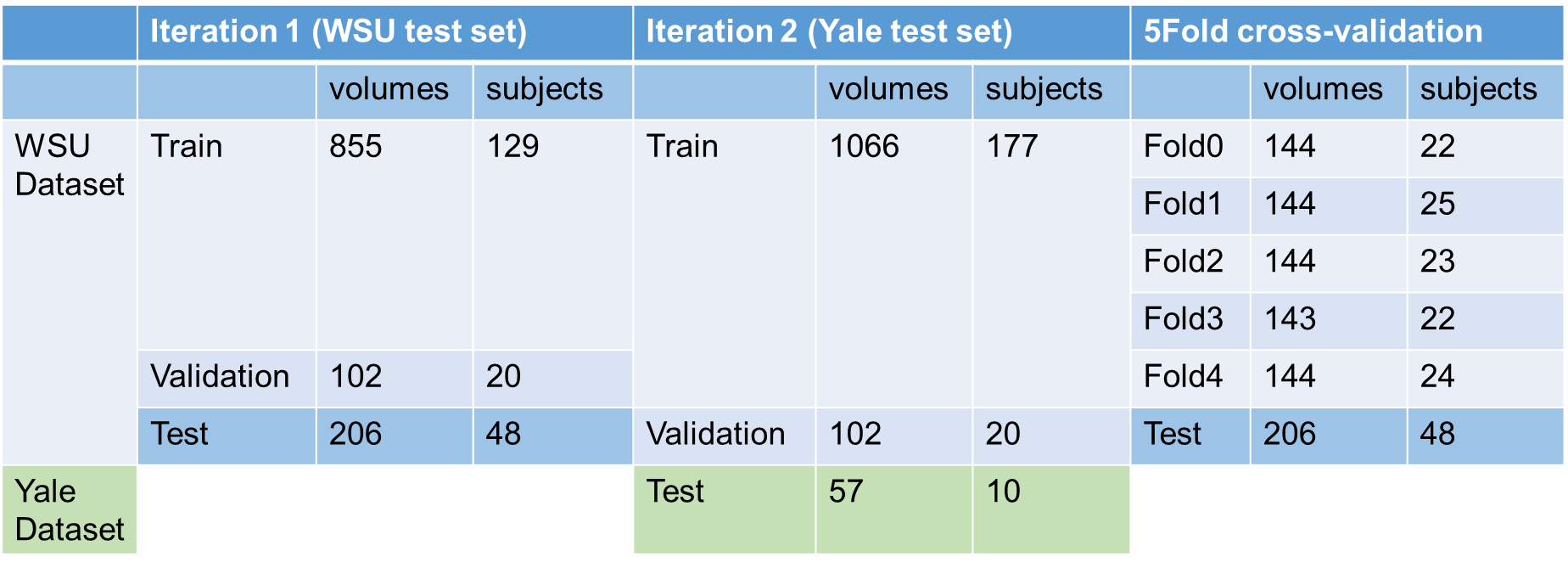}

Three types of AI models were employed: 3D UNet, VNet, and HighRes-net.

We used our model based on the following references.

\begin{enumerate}
\def\labelenumi{\arabic{enumi}.}
\item
  3D UNet: This is an improved version of UNet, enhanced with residual
  units using the ResidualUnit class. The residual part uses convolution
  to adjust input dimensions to match the output dimensions when needed,
  or nn.Identity if not. (link\footnote{https://docs.monai.io/en/stable/networks.html\#unet},
  and UNet Research Paper \footnote{https://link.springer.com/chapter/10.1007/978-3-030-12029-0\_40})
  (13)
\item
  VNet: This is a V-Net model based on Fully Convolutional Neural
  Networks designed for volumetric medical image segmentation.
  It\textquotesingle s adapted from the official Caffe implementation
  \footnote{https://github.com/faustomilletari/VNet}, and another
  PyTorch implementation \footnote{\href{https://github.com/mattmacy/vnet.pytorch/blob/master/vnet.py}{vnet.pytorch/vnet.py
    at master · mattmacy/vnet.pytorch (github.com)}}. The model is
  versatile and can handle both 2D and 3D inputs
\item
  HighRes-Net: This is a reimplementation of the highres3dnet model
  inspired by Li et al.\textquotesingle s work titled "On the
  compactness, efficiency, and representation of 3D convolutional
  networks: Brain parcellation as a pretext task", presented at IPMI
  \textquotesingle17. (link \footnote{\href{https://docs.monai.io/en/stable/networks.html\#highresnet}{Network
    architectures --- MONAI 1.2.0 Documentation}}) (14)
\end{enumerate}

\includegraphics[width=6.5in,height=4.24722in]{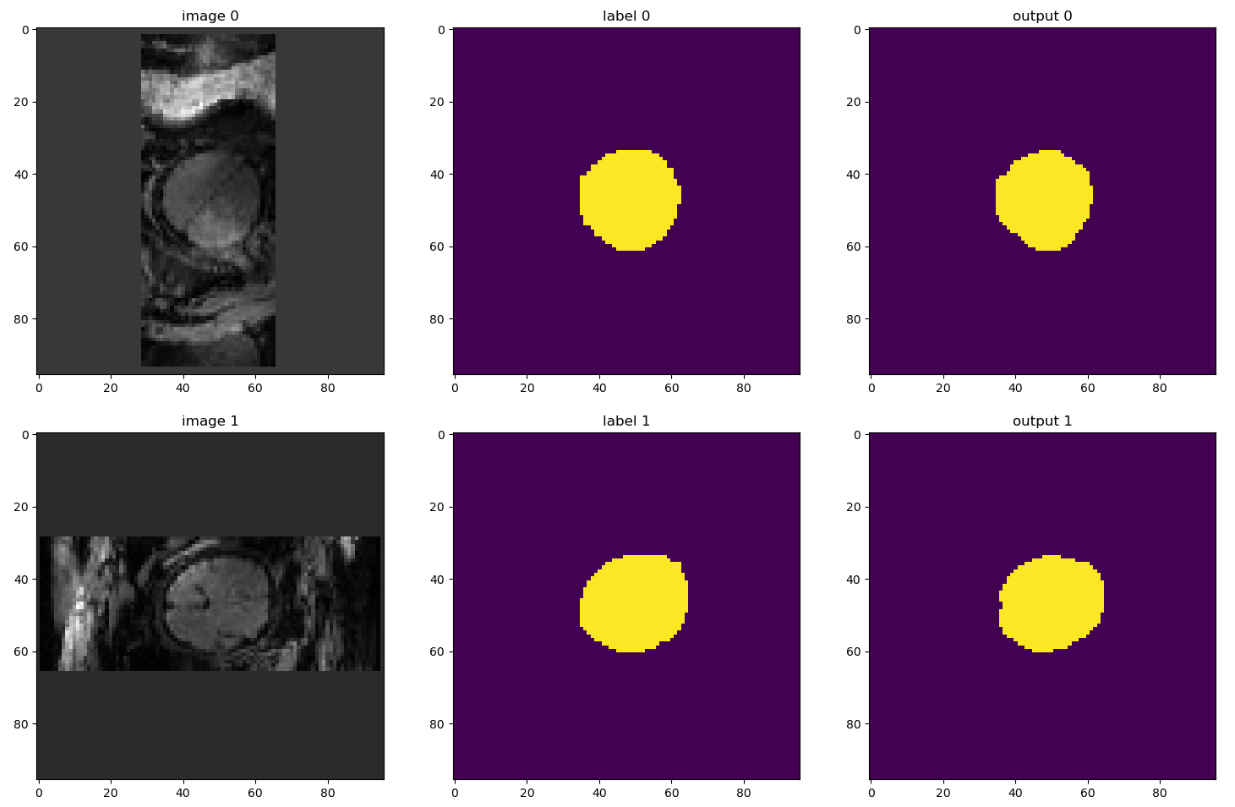}

Figure 1- Left: Original fMRI; Middle: Brain Mask (provided by dataset);
Right: AI-predicated Brain Mask, showing good accuracy

\hypertarget{this-is-how-we-divided-it-into-five-folders}{%
\section{This is how we divided it into five
folders:}\label{this-is-how-we-divided-it-into-five-folders}}

The provided table outlines the division of data into five folds to
perform 5-fold cross-validation and the statistics for the test dataset.
The table provides information regarding the number of 3D volumes, 4D
fMRI sets, and patients in each fold.

\begin{enumerate}
\def\labelenumi{\arabic{enumi}.}
\item
  5-Fold Cross-Validation:

  \begin{itemize}
  \item
    The data were divided into five folds, labeled Fold0 to Fold4.
  \item
    Each fold represents a subset of the dataset used for the training
    and validation.
  \item
    The number of 3D volumes, 4D fMRI sets, and patients differed
    slightly across the folds, indicating some variability in dataset
    composition.
  \end{itemize}
\item
  Test Dataset:

  \begin{itemize}
  \item
    This table also provides information on the test dataset.
  \item
    The test dataset is stated to be the same as the original paper,
  \end{itemize}
\end{enumerate}

\begin{quote}
This suggests that this is an external dataset used for evaluation
purposes. We followed the test division of the original paper to respect
the original author and facilitate the comparison.
\end{quote}

\begin{itemize}
\item
  The test dataset contains 206 3D volumes, 48 of 4D fMRI sets, and 37
  patients.
\end{itemize}

\begin{longtable}[]{@{}
  >{\raggedright\arraybackslash}p{(\columnwidth - 6\tabcolsep) * \real{0.1924}}
  >{\raggedright\arraybackslash}p{(\columnwidth - 6\tabcolsep) * \real{0.2770}}
  >{\raggedright\arraybackslash}p{(\columnwidth - 6\tabcolsep) * \real{0.2675}}
  >{\raggedright\arraybackslash}p{(\columnwidth - 6\tabcolsep) * \real{0.2632}}@{}}
\caption{Table 1- Dataset distribution}\tabularnewline
\toprule()
\begin{minipage}[b]{\linewidth}\raggedright
\textbf{~}
\end{minipage} & \begin{minipage}[b]{\linewidth}\raggedright
\textbf{Volumes}
\end{minipage} & \begin{minipage}[b]{\linewidth}\raggedright
\textbf{Subjects}
\end{minipage} & \begin{minipage}[b]{\linewidth}\raggedright
\textbf{Patients}
\end{minipage} \\
\midrule()
\endfirsthead
\toprule()
\begin{minipage}[b]{\linewidth}\raggedright
\textbf{~}
\end{minipage} & \begin{minipage}[b]{\linewidth}\raggedright
\textbf{Volumes}
\end{minipage} & \begin{minipage}[b]{\linewidth}\raggedright
\textbf{Subjects}
\end{minipage} & \begin{minipage}[b]{\linewidth}\raggedright
\textbf{Patients}
\end{minipage} \\
\midrule()
\endhead
\textbf{Fold0} & 144 & 22 & 21 \\
\textbf{Fold1} & 144 & 25 & 21 \\
\textbf{Fold2} & 144 & 23 & 21 \\
\textbf{Fold3} & 143 & 22 & 21 \\
\textbf{Fold4} & 144 & 24 & 23 \\
\textbf{Test} & 206 & 48 & 48 \\
\bottomrule()
\end{longtable}

\includegraphics[width=5.98472in,height=5.2071in]{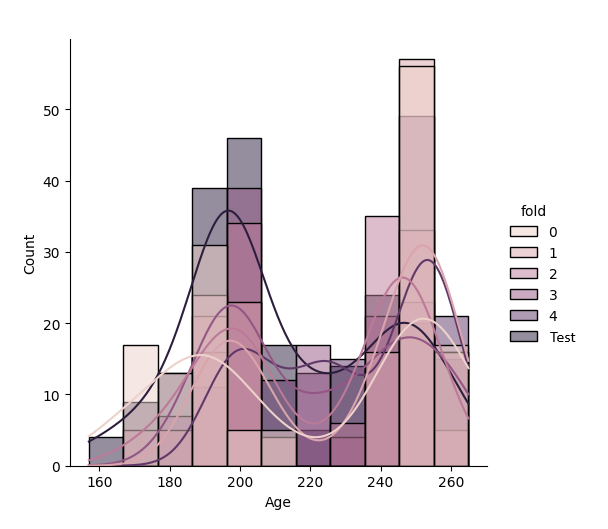}

Figure 2- Distribution of original dataset according to GA days

\hypertarget{cross-validation}{%
\section{Cross-Validation:}\label{cross-validation}}

This section highlights details regarding the data distribution strategy
as well as precautions taken during the division into folds to provide
transparency and ensure the reliability of the cross-validation results.

The downloaded dataset comprises 155 subdirectories representing
individual patients. Additional subdirectories contained fMRI images at
different gestational ages within these patient directories. The dataset
consisted of 164 participants. Each 4D fMRI image within the dataset
contains multiple 3D image volumes, resulting in a total of 925 volumes
after decomposition.

It is important to note that the number of 3D volumes in fMRI data was
inconsistent across the dataset. To address this issue during data
division into folds, we took precautions to ensure that 3D volumes from
the same fMRI were not present in different folds. This approach helps
prevent data leakage and ensures the integrity of the cross-validation
process.

For example, in Fold0, which is comprised of 21 patients, we included a
total of 22 fMRI files. Consequently, the division resulted in 144 3D
volumes for the training and validation.

By carefully managing the distribution of 3D volumes from the same fMRI
within the folds, we aimed to maintain the integrity of the
cross-validation process and ensure that the models were trained and
evaluated on independent data subsets. This approach allows for a robust
assessment of model performance and generalization capabilities.

\textbf{Resources and Tutorial}

To ensure the reproducibility and applicability of our research, we have
made our code, protocols, and tutorial Python notebook available on
GitHub(https://github.com/Achillesy/Fetal\_Functional\_MRI\_Segmentation).
This repository provides researchers and practitioners with the
necessary resources to apply our model to their own data and contexts,
further promoting the use of AI in fetal brain fMRI segmentation.

In addition, raw volumes and manually drawn brain masks used for
training and validating our model were hosted on OpenNeuro.org. Other
researchers can use these resources to further develop and validate AI
models for fetal brain segmentation.

\hypertarget{optuna}{%
\section{\texorpdfstring{Optuna: }{Optuna: }}\label{optuna}}

Optuna was used to automatically select hyperparameters.

Optuna is a software framework for automating the optimization process
of these hyperparameters. It automatically finds optimal hyperparameter
values using different samplers, such as grid search, random, Bayesian,
and evolutionary algorithms. Herein, we describe the different samplers
available in Optuna.

\begin{itemize}
\item
  Grid Search: The search space for each hyperparameter is discretized.
  The optimizer launches learning for each hyperparameter configuration,
  and selects the best at the end.
\item
  Random: Randomly samples the search space and continues until the
  stopping criteria are satisfied.
\item
  \textbf{Bayesian: Probabilistic model-based approach for finding
  optimal hyperparameters. (This was our main sampler in this project)}
\item
  Evolutionary algorithms: Meta-heuristic approaches employ the value of
  the fitness function to determine the optimal hyperparameters.
\end{itemize}

In this study, we applied the following code to define the Optuna
hyperparameter range:

\# Optimize

cfg.lr = trial.suggest\_loguniform("lr", 1e-4, 1e-3)

cfg.weight\_decay = trial.suggest\_loguniform("weight\_decay", 1e-5,
1e-4)

optimizer = torch.optim.Adam(model.parameters(), lr=cfg.lr,
~weight\_decay=cfg.weight\_decay)

\# Learning schedule

cfg.step\_size = trial.suggest\_int("step\_size", 5, 20)

cfg.factor = trial.suggest\_loguniform("factor", 0.1, 0.9)

scheduler = torch.optim.lr\_scheduler.StepLR(

~ ~ optimizer, step\_size=cfg.step\_size, gamma=cfg.factor

)

\# model dropout\_reate

cfg.dropout\_rate = trial.suggest\_uniform("dropout\_rate", 0.0, 0.5)

\# loss function

loss\_function = DiceLoss(to\_onehot\_y=True, softmax=True)

dice\_metric = DiceMetric(include\_background=False, reduction="mean")

\hypertarget{results}{%
\section{Results}\label{results}}

\hypertarget{result-1-general-results.}{%
\section{Result 1: General results.}\label{result-1-general-results.}}

\textbf{Auto-masking Performance}

Our three AI models (3D UNet, VNet, and HighRes-Net) demonstrated
commendable performance in fetal brain fMRI segmentation. We evaluated
these models using a 5-fold cross-validation approach on an open-source
dataset of fetal fMRI comprising 155 cases(patients). The table below
shows the results.

\textbf{Computational Time and Hardware}

Our models demonstrate the advantage of deep learning in terms of speed.
For instance, the VNet model, which was trained on a GeForce GTX 1080 Ti
GPU, took about 5 hours to train.

We provide a test program that can be run directly in Colab
(https://colab.research.google.com/github/Achillesy/Fetal\_Functional\_MRI\_Segmentation/blob/master/fmri\_vnet\_interface.ipynb),
which can now generate multiple corresponding masks for fMRI files
(including numerous 3D images). The generation time depends on the
number of 3D images included in the fMRI. With the GPU turned on (free
users), the average build time is about 1.2 seconds and about 17 seconds
without hardware acceleration.

\hypertarget{results-2-comparison-of-architecture-of-each-model}{%
\section{Results 2: Comparison of architecture of each
model}\label{results-2-comparison-of-architecture-of-each-model}}

\includegraphics[width=2.20134in,height=2.66092in]{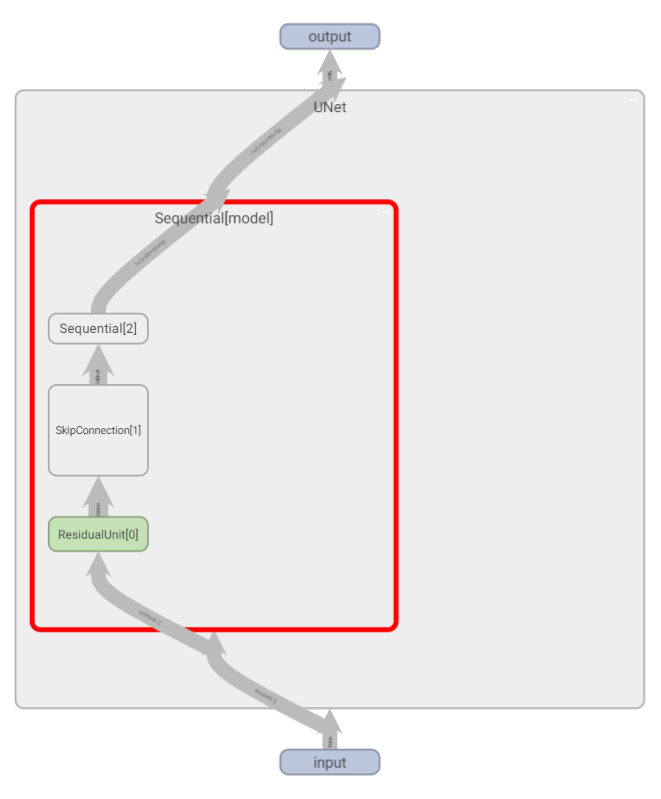}
\includegraphics[width=2.30591in,height=2.81609in]{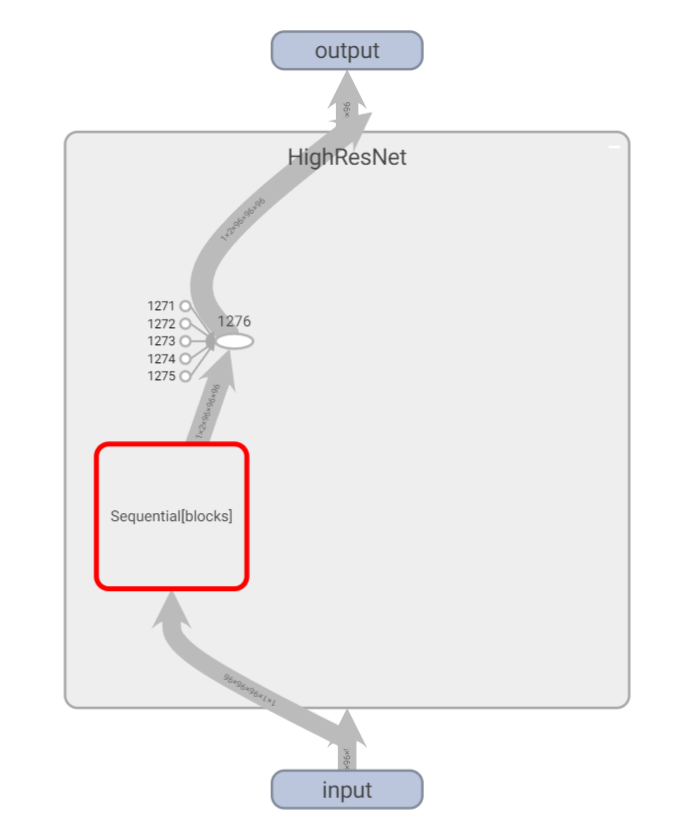}
\includegraphics[width=1.85632in,height=3.29335in]{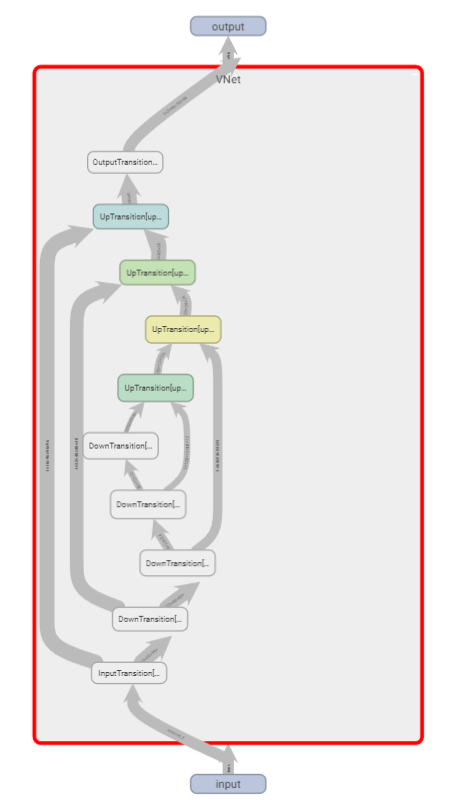}

Figure 4- Schematics of three different models: Left: UNet; Middle:
HighResNet; Right: VNet

\begin{longtable}[]{@{}
  >{\raggedright\arraybackslash}p{(\columnwidth - 6\tabcolsep) * \real{0.2500}}
  >{\raggedright\arraybackslash}p{(\columnwidth - 6\tabcolsep) * \real{0.2500}}
  >{\raggedright\arraybackslash}p{(\columnwidth - 6\tabcolsep) * \real{0.2500}}
  >{\raggedright\arraybackslash}p{(\columnwidth - 6\tabcolsep) * \real{0.2500}}@{}}
\caption{Table 2- Distribution of dataset into 5- folders for cross
validation and test}\tabularnewline
\toprule()
\begin{minipage}[b]{\linewidth}\raggedright
\textbf{Parameter}
\end{minipage} & \begin{minipage}[b]{\linewidth}\raggedright
\textbf{UNet}
\end{minipage} & \begin{minipage}[b]{\linewidth}\raggedright
\textbf{HighResNet}
\end{minipage} & \begin{minipage}[b]{\linewidth}\raggedright
\textbf{VNet}
\end{minipage} \\
\midrule()
\endfirsthead
\toprule()
\begin{minipage}[b]{\linewidth}\raggedright
\textbf{Parameter}
\end{minipage} & \begin{minipage}[b]{\linewidth}\raggedright
\textbf{UNet}
\end{minipage} & \begin{minipage}[b]{\linewidth}\raggedright
\textbf{HighResNet}
\end{minipage} & \begin{minipage}[b]{\linewidth}\raggedright
\textbf{VNet}
\end{minipage} \\
\midrule()
\endhead
Architecture & Encoder-decoder with skip connections & Encoder-decoder
with residual connections & 3D convolutional network \\
Dataset &
\multicolumn{3}{>{\raggedright\arraybackslash}p{(\columnwidth - 6\tabcolsep) * \real{0.7500} + 4\tabcolsep}@{}}{%
Varies based on the application} \\
Evaluation Metrics &
\multicolumn{3}{>{\raggedright\arraybackslash}p{(\columnwidth - 6\tabcolsep) * \real{0.7500} + 4\tabcolsep}@{}}{%
Dice Score , HD sensitivity,} \\
Training Strategy & Adam, learning rate, decay, steplr scheduler, and
early stopping & Adam, learning rate, decay, steplr scheduler, and early
stopping & Adam, learning rate, decay, steplr scheduler, and early
stopping \\
Computational Resources &
\multicolumn{3}{>{\raggedright\arraybackslash}p{(\columnwidth - 6\tabcolsep) * \real{0.7500} + 4\tabcolsep}@{}}{%
GPU with sufficient memory} \\
Pretrained Weights & No & No & No \\
Regularization Techniques & Batch normalization, dropout, weight decay &
Batch normalization, dropout,weight decay & Batch normalization,
dropout, weight decay \\
Comparative Analysis & Achieves high accuracy in medical image
segmentation tasks & Achieves high accuracy in medical image
segmentation tasks, lower memory requirements than UNET & Achieves high
accuracy in medical image segmentation tasks, better performance on
small datasets \\
Qualitative Results & Accurately segments organs and tissues in medical
images & Accurately segments organs and tissues in medical images &
Accurately segments organs and tissues in medical images, can handle 3D
volumes \\
\bottomrule()
\end{longtable}

\hypertarget{result-3-performance-of-each-model.}{%
\section{Result 3: Performance of each
model.}\label{result-3-performance-of-each-model.}}

\hypertarget{result-3.1-performance-for-the-unet-model}{%
\section{Result 3.1: Performance for the Unet
Model}\label{result-3.1-performance-for-the-unet-model}}

In this section, we discuss the performance of the UNet model for fetal
resting-state functional MRI brain segmentation. The table consists of
two sections: one for the manually selected hyperparameters and the
other for the hyperparameters selected by Optuna, an automated
hyperparameter-tuning tool. The metrics used for evaluation included the
Dice scores for the validation and test sets across five folds.

The UNet model achieved competitive results for the manually selected
hyperparameters. The Dice scores for the validation set ranged from
0.9114 to 0.9408, whereas for the test set, they ranged from 0.9035 to
0.9260. These scores indicated the accuracy of the model in segmenting
the fetal brain from the surrounding tissue.

When Optuna was employed to automatically select hyperparameters, the
UNet model showed further improvement. The Dice scores for the
validation set ranged from 0.9181 to 0.9440; for the test set, they
ranged from 0.9060 to 0.9282. This demonstrates the effectiveness of the
hyperparameter optimization in enhancing the performance of the model.

Above: Manually selected hyperparameters

Optuna-Selected Hyperparameters

\begin{longtable}[]{@{}
  >{\raggedright\arraybackslash}p{(\columnwidth - 10\tabcolsep) * \real{0.0918}}
  >{\raggedright\arraybackslash}p{(\columnwidth - 10\tabcolsep) * \real{0.3198}}
  >{\raggedright\arraybackslash}p{(\columnwidth - 10\tabcolsep) * \real{0.1406}}
  >{\raggedright\arraybackslash}p{(\columnwidth - 10\tabcolsep) * \real{0.1406}}
  >{\raggedright\arraybackslash}p{(\columnwidth - 10\tabcolsep) * \real{0.0946}}
  >{\raggedright\arraybackslash}p{(\columnwidth - 10\tabcolsep) * \real{0.2126}}@{}}
\caption{Table 3- Comparison of architecture for three different
models}\tabularnewline
\toprule()
\multicolumn{6}{@{}>{\raggedright\arraybackslash}p{(\columnwidth - 10\tabcolsep) * \real{1.0000} + 10\tabcolsep}@{}}{%
\begin{minipage}[b]{\linewidth}\raggedright
UNet Model: Manually-selected hyperparameters:
\end{minipage}} \\
\midrule()
\endfirsthead
\toprule()
\multicolumn{6}{@{}>{\raggedright\arraybackslash}p{(\columnwidth - 10\tabcolsep) * \real{1.0000} + 10\tabcolsep}@{}}{%
\begin{minipage}[b]{\linewidth}\raggedright
UNet Model: Manually-selected hyperparameters:
\end{minipage}} \\
\midrule()
\endhead
~ & \textbf{Training} & \textbf{Validate} & \textbf{Dice of Validate} &
\textbf{Test} & \textbf{Dice of Test} \\
Fold 0 & 575 & 144 & 0.9310 & 206 & 0.9260±0.0960 \\
Fold 1 & 575 & 144 & 0.9408 & 206 & 0.9108±0.1296 \\
Fold 2 & 575 & 144 & 0.9301 & 206 & 0.9025±0.1303 \\
Fold 3 & 576 & 143 & 0.9114 & 206 & 0.9097±0.1034 \\
Fold 4 & 575 & 144 & 0.9311 & 206 & 0.9035±0.1065 \\
\multicolumn{6}{@{}>{\raggedright\arraybackslash}p{(\columnwidth - 10\tabcolsep) * \real{1.0000} + 10\tabcolsep}@{}}{%
UNet Model: Optuna-selected hyperparameters:} \\
~ & \textbf{Training} & \textbf{Validate} & \textbf{Dice of Validate} &
\textbf{Test} & \textbf{Dice of Test} \\
Fold 0 & 575 & 144 & 0.9371 & 206 & 0.9282±0.0768 \\
Fold 1 & 575 & 144 & 0.9440 & 206 & 0.9260±0.1140 \\
Fold 2 & 575 & 144 & 0.9380 & 206 & 0.9228±0.1025 \\
Fold 3 & 576 & 143 & 0.9181 & 206 & 0.9060±0.1018 \\
Fold 4 & 575 & 144 & 0.9403 & 206 & 0.9219±0.1165 \\
\bottomrule()
\end{longtable}

It can be inferred that both the manually selected and Optuna-selected
hyperparameters yielded a good segmentation performance with the UNet
model. However, the Optuna-selected hyperparameters led to slightly
higher Dice scores, indicating improved accuracy.

\hypertarget{result-3.2-performance-for-the-vnet-model}{%
\section{Result 3.2: Performance for the VNet
Model}\label{result-3.2-performance-for-the-vnet-model}}

By analyzing the provided table for the VNet model, we can observe the
performance of this model for fetal resting-state fMRI brain
segmentation. Similar to the UNet model, the table consists of two
sections: one for the manually selected hyperparameters, and the other
for the hyperparameters selected by Optuna.

The VNet model achieved competitive results for the manually selected
hyperparameters. The dice scores for the validation set ranged from
0.9254 to 0.9521, while for the test set, they ranged from 0.9126 to
0.9355. These scores indicated the accuracy of the model in segmenting
the fetal brain from the surrounding tissue.

When Optuna was employed to automatically select hyperparameters, the
VNet model showed improved performance. The dice scores for the
validation set ranged from 0.9277 to 0.9493, and for the test set, they
ranged from 0.9181 to 0.9286. This demonstrates that the automatic
hyperparameter optimization led to slightly higher Dice scores,
indicating enhanced accuracy compared to the manually selected
hyperparameters.

Above: Manually selected hyperparameters

Optuna-Selected Hyperparameters

\begin{longtable}[]{@{}
  >{\raggedright\arraybackslash}p{(\columnwidth - 20\tabcolsep) * \real{0.0915}}
  >{\raggedright\arraybackslash}p{(\columnwidth - 20\tabcolsep) * \real{0.0169}}
  >{\raggedright\arraybackslash}p{(\columnwidth - 20\tabcolsep) * \real{0.2346}}
  >{\raggedright\arraybackslash}p{(\columnwidth - 20\tabcolsep) * \real{0.0682}}
  >{\raggedright\arraybackslash}p{(\columnwidth - 20\tabcolsep) * \real{0.0896}}
  >{\raggedright\arraybackslash}p{(\columnwidth - 20\tabcolsep) * \real{0.0510}}
  >{\raggedright\arraybackslash}p{(\columnwidth - 20\tabcolsep) * \real{0.1069}}
  >{\raggedright\arraybackslash}p{(\columnwidth - 20\tabcolsep) * \real{0.0338}}
  >{\raggedright\arraybackslash}p{(\columnwidth - 20\tabcolsep) * \real{0.0778}}
  >{\raggedright\arraybackslash}p{(\columnwidth - 20\tabcolsep) * \real{0.0168}}
  >{\raggedright\arraybackslash}p{(\columnwidth - 20\tabcolsep) * \real{0.2128}}@{}}
\caption{Table 4 Results for the UNet Model in fMRI brain
extraction}\tabularnewline
\toprule()
\multicolumn{11}{@{}>{\raggedright\arraybackslash}p{(\columnwidth - 20\tabcolsep) * \real{1.0000} + 20\tabcolsep}@{}}{%
\begin{minipage}[b]{\linewidth}\raggedright
\begin{quote}
VNet Model: Manually-selected hyperparameters:
\end{quote}
\end{minipage}} \\
\midrule()
\endfirsthead
\toprule()
\multicolumn{11}{@{}>{\raggedright\arraybackslash}p{(\columnwidth - 20\tabcolsep) * \real{1.0000} + 20\tabcolsep}@{}}{%
\begin{minipage}[b]{\linewidth}\raggedright
\begin{quote}
VNet Model: Manually-selected hyperparameters:
\end{quote}
\end{minipage}} \\
\midrule()
\endhead
~ &
\multicolumn{3}{>{\raggedright\arraybackslash}p{(\columnwidth - 20\tabcolsep) * \real{0.3197} + 4\tabcolsep}}{%
\textbf{Training}} &
\multicolumn{2}{>{\raggedright\arraybackslash}p{(\columnwidth - 20\tabcolsep) * \real{0.1406} + 2\tabcolsep}}{%
\textbf{Validate}} &
\multicolumn{2}{>{\raggedright\arraybackslash}p{(\columnwidth - 20\tabcolsep) * \real{0.1407} + 2\tabcolsep}}{%
\textbf{Dice of Validate}} &
\multicolumn{2}{>{\raggedright\arraybackslash}p{(\columnwidth - 20\tabcolsep) * \real{0.0946} + 2\tabcolsep}}{%
\textbf{Test}} & \textbf{Dice of Test} \\
Fold 0 &
\multicolumn{3}{>{\raggedright\arraybackslash}p{(\columnwidth - 20\tabcolsep) * \real{0.3197} + 4\tabcolsep}}{%
575} &
\multicolumn{2}{>{\raggedright\arraybackslash}p{(\columnwidth - 20\tabcolsep) * \real{0.1406} + 2\tabcolsep}}{%
144} &
\multicolumn{2}{>{\raggedright\arraybackslash}p{(\columnwidth - 20\tabcolsep) * \real{0.1407} + 2\tabcolsep}}{%
0.9254} &
\multicolumn{2}{>{\raggedright\arraybackslash}p{(\columnwidth - 20\tabcolsep) * \real{0.0946} + 2\tabcolsep}}{%
206} & 0.9215±0.1340 \\
Fold 1 &
\multicolumn{3}{>{\raggedright\arraybackslash}p{(\columnwidth - 20\tabcolsep) * \real{0.3197} + 4\tabcolsep}}{%
575} &
\multicolumn{2}{>{\raggedright\arraybackslash}p{(\columnwidth - 20\tabcolsep) * \real{0.1406} + 2\tabcolsep}}{%
144} &
\multicolumn{2}{>{\raggedright\arraybackslash}p{(\columnwidth - 20\tabcolsep) * \real{0.1407} + 2\tabcolsep}}{%
0.9485} &
\multicolumn{2}{>{\raggedright\arraybackslash}p{(\columnwidth - 20\tabcolsep) * \real{0.0946} + 2\tabcolsep}}{%
206} & 0.9126±0.1522 \\
Fold 2 &
\multicolumn{3}{>{\raggedright\arraybackslash}p{(\columnwidth - 20\tabcolsep) * \real{0.3197} + 4\tabcolsep}}{%
575} &
\multicolumn{2}{>{\raggedright\arraybackslash}p{(\columnwidth - 20\tabcolsep) * \real{0.1406} + 2\tabcolsep}}{%
144} &
\multicolumn{2}{>{\raggedright\arraybackslash}p{(\columnwidth - 20\tabcolsep) * \real{0.1407} + 2\tabcolsep}}{%
0.9493} &
\multicolumn{2}{>{\raggedright\arraybackslash}p{(\columnwidth - 20\tabcolsep) * \real{0.0946} + 2\tabcolsep}}{%
206} & 0.9131±0.1528 \\
Fold 3 &
\multicolumn{3}{>{\raggedright\arraybackslash}p{(\columnwidth - 20\tabcolsep) * \real{0.3197} + 4\tabcolsep}}{%
576} &
\multicolumn{2}{>{\raggedright\arraybackslash}p{(\columnwidth - 20\tabcolsep) * \real{0.1406} + 2\tabcolsep}}{%
143} &
\multicolumn{2}{>{\raggedright\arraybackslash}p{(\columnwidth - 20\tabcolsep) * \real{0.1407} + 2\tabcolsep}}{%
0.9408} &
\multicolumn{2}{>{\raggedright\arraybackslash}p{(\columnwidth - 20\tabcolsep) * \real{0.0946} + 2\tabcolsep}}{%
206} & 0.9300±0.0829 \\
Fold 4 &
\multicolumn{3}{>{\raggedright\arraybackslash}p{(\columnwidth - 20\tabcolsep) * \real{0.3197} + 4\tabcolsep}}{%
575} &
\multicolumn{2}{>{\raggedright\arraybackslash}p{(\columnwidth - 20\tabcolsep) * \real{0.1406} + 2\tabcolsep}}{%
144} &
\multicolumn{2}{>{\raggedright\arraybackslash}p{(\columnwidth - 20\tabcolsep) * \real{0.1407} + 2\tabcolsep}}{%
0.9521} &
\multicolumn{2}{>{\raggedright\arraybackslash}p{(\columnwidth - 20\tabcolsep) * \real{0.0946} + 2\tabcolsep}}{%
206} & 0.9355±0.0757 \\
\multicolumn{11}{@{}>{\raggedright\arraybackslash}p{(\columnwidth - 20\tabcolsep) * \real{1.0000} + 20\tabcolsep}@{}}{%
VNet Model: Optuna-selected hyperparameters} \\
\multicolumn{2}{@{}>{\raggedright\arraybackslash}p{(\columnwidth - 20\tabcolsep) * \real{0.1085} + 2\tabcolsep}}{%
~} & \textbf{Training} &
\multicolumn{2}{>{\raggedright\arraybackslash}p{(\columnwidth - 20\tabcolsep) * \real{0.1578} + 2\tabcolsep}}{%
\textbf{Validate}} &
\multicolumn{2}{>{\raggedright\arraybackslash}p{(\columnwidth - 20\tabcolsep) * \real{0.1578} + 2\tabcolsep}}{%
\textbf{Dice of Validate}} &
\multicolumn{2}{>{\raggedright\arraybackslash}p{(\columnwidth - 20\tabcolsep) * \real{0.1117} + 2\tabcolsep}}{%
\textbf{Test}} &
\multicolumn{2}{>{\raggedright\arraybackslash}p{(\columnwidth - 20\tabcolsep) * \real{0.2297} + 2\tabcolsep}@{}}{%
\textbf{Dice of Test}} \\
\multicolumn{2}{@{}>{\raggedright\arraybackslash}p{(\columnwidth - 20\tabcolsep) * \real{0.1085} + 2\tabcolsep}}{%
Fold 0} & 575 &
\multicolumn{2}{>{\raggedright\arraybackslash}p{(\columnwidth - 20\tabcolsep) * \real{0.1578} + 2\tabcolsep}}{%
144} &
\multicolumn{2}{>{\raggedright\arraybackslash}p{(\columnwidth - 20\tabcolsep) * \real{0.1578} + 2\tabcolsep}}{%
0.9277} &
\multicolumn{2}{>{\raggedright\arraybackslash}p{(\columnwidth - 20\tabcolsep) * \real{0.1117} + 2\tabcolsep}}{%
206} &
\multicolumn{2}{>{\raggedright\arraybackslash}p{(\columnwidth - 20\tabcolsep) * \real{0.2297} + 2\tabcolsep}@{}}{%
0.9277±0.0839} \\
\multicolumn{2}{@{}>{\raggedright\arraybackslash}p{(\columnwidth - 20\tabcolsep) * \real{0.1085} + 2\tabcolsep}}{%
Fold 1} & 575 &
\multicolumn{2}{>{\raggedright\arraybackslash}p{(\columnwidth - 20\tabcolsep) * \real{0.1578} + 2\tabcolsep}}{%
144} &
\multicolumn{2}{>{\raggedright\arraybackslash}p{(\columnwidth - 20\tabcolsep) * \real{0.1578} + 2\tabcolsep}}{%
0.9490} &
\multicolumn{2}{>{\raggedright\arraybackslash}p{(\columnwidth - 20\tabcolsep) * \real{0.1117} + 2\tabcolsep}}{%
206} &
\multicolumn{2}{>{\raggedright\arraybackslash}p{(\columnwidth - 20\tabcolsep) * \real{0.2297} + 2\tabcolsep}@{}}{%
0.9286±0.0812} \\
\multicolumn{2}{@{}>{\raggedright\arraybackslash}p{(\columnwidth - 20\tabcolsep) * \real{0.1085} + 2\tabcolsep}}{%
Fold 2} & 575 &
\multicolumn{2}{>{\raggedright\arraybackslash}p{(\columnwidth - 20\tabcolsep) * \real{0.1578} + 2\tabcolsep}}{%
144} &
\multicolumn{2}{>{\raggedright\arraybackslash}p{(\columnwidth - 20\tabcolsep) * \real{0.1578} + 2\tabcolsep}}{%
0.9471} &
\multicolumn{2}{>{\raggedright\arraybackslash}p{(\columnwidth - 20\tabcolsep) * \real{0.1117} + 2\tabcolsep}}{%
206} &
\multicolumn{2}{>{\raggedright\arraybackslash}p{(\columnwidth - 20\tabcolsep) * \real{0.2297} + 2\tabcolsep}@{}}{%
0.9189±0.1237} \\
\multicolumn{2}{@{}>{\raggedright\arraybackslash}p{(\columnwidth - 20\tabcolsep) * \real{0.1085} + 2\tabcolsep}}{%
Fold 3} & 576 &
\multicolumn{2}{>{\raggedright\arraybackslash}p{(\columnwidth - 20\tabcolsep) * \real{0.1578} + 2\tabcolsep}}{%
143} &
\multicolumn{2}{>{\raggedright\arraybackslash}p{(\columnwidth - 20\tabcolsep) * \real{0.1578} + 2\tabcolsep}}{%
0.9388} &
\multicolumn{2}{>{\raggedright\arraybackslash}p{(\columnwidth - 20\tabcolsep) * \real{0.1117} + 2\tabcolsep}}{%
206} &
\multicolumn{2}{>{\raggedright\arraybackslash}p{(\columnwidth - 20\tabcolsep) * \real{0.2297} + 2\tabcolsep}@{}}{%
0.9226±0.0886} \\
\multicolumn{2}{@{}>{\raggedright\arraybackslash}p{(\columnwidth - 20\tabcolsep) * \real{0.1085} + 2\tabcolsep}}{%
Fold 4} & 575 &
\multicolumn{2}{>{\raggedright\arraybackslash}p{(\columnwidth - 20\tabcolsep) * \real{0.1578} + 2\tabcolsep}}{%
144} &
\multicolumn{2}{>{\raggedright\arraybackslash}p{(\columnwidth - 20\tabcolsep) * \real{0.1578} + 2\tabcolsep}}{%
0.9479} &
\multicolumn{2}{>{\raggedright\arraybackslash}p{(\columnwidth - 20\tabcolsep) * \real{0.1117} + 2\tabcolsep}}{%
206} &
\multicolumn{2}{>{\raggedright\arraybackslash}p{(\columnwidth - 20\tabcolsep) * \real{0.2297} + 2\tabcolsep}@{}}{%
0.9181±0.1018} \\
\bottomrule()
\end{longtable}

Based on these results, it can be inferred that both the manually
selected and Optuna-selected hyperparameters yielded a good segmentation
performance with the VNet model. However, the Optuna-selected
hyperparameters led to slightly higher Dice scores, indicating improved
accuracy.

\hypertarget{result-3.3-performance-for-highres-net-model}{%
\section{Result 3.3: Performance for HighRes-Net
Model}\label{result-3.3-performance-for-highres-net-model}}

By analyzing the provided table for the HighRes-Net model, we can
observe the performance of this model for fetal resting-state fMRI brain
segmentation. Similar to the previous tables, this table also consists
of two sections: one for the manually selected hyperparameters, and the
other for the hyperparameters selected by Optuna.

The HighRes-Net model achieved varied results for the manually selected
hyperparameters. The Dice scores for the validation set ranged from
0.9095 to 0.9459, whereas for the test set, they ranged from 0.9038 to
0.9138. These scores indicated the accuracy of the model in segmenting
the fetal brain from the surrounding tissue.

When Optuna was employed to select hyperparameters automatically, the
HighRes-Net model still exhibited mixed performance. The dice scores for
the validation set ranged from 0.9142 to 0.9374; for the test set, they
ranged from 0.8803 to 0.9127. Notably, the dice scores obtained from the
Optuna hyperparameters were generally lower than those obtained using
manually selected hyperparameters.

Above: Manually selected hyperparameters

Optuna-Selected Hyperparameters

\begin{longtable}[]{@{}
  >{\raggedright\arraybackslash}p{(\columnwidth - 10\tabcolsep) * \real{0.1088}}
  >{\raggedright\arraybackslash}p{(\columnwidth - 10\tabcolsep) * \real{0.2346}}
  >{\raggedright\arraybackslash}p{(\columnwidth - 10\tabcolsep) * \real{0.1578}}
  >{\raggedright\arraybackslash}p{(\columnwidth - 10\tabcolsep) * \real{0.1578}}
  >{\raggedright\arraybackslash}p{(\columnwidth - 10\tabcolsep) * \real{0.1116}}
  >{\raggedright\arraybackslash}p{(\columnwidth - 10\tabcolsep) * \real{0.2294}}@{}}
\caption{Table 5 Results for the VNet Model in fMRI brain
extraction}\tabularnewline
\toprule()
\multicolumn{6}{@{}>{\raggedright\arraybackslash}p{(\columnwidth - 10\tabcolsep) * \real{1.0000} + 10\tabcolsep}@{}}{%
\begin{minipage}[b]{\linewidth}\raggedright
HighRes-Net Model: Manually-selected hyperparameters
\end{minipage}} \\
\midrule()
\endfirsthead
\toprule()
\multicolumn{6}{@{}>{\raggedright\arraybackslash}p{(\columnwidth - 10\tabcolsep) * \real{1.0000} + 10\tabcolsep}@{}}{%
\begin{minipage}[b]{\linewidth}\raggedright
HighRes-Net Model: Manually-selected hyperparameters
\end{minipage}} \\
\midrule()
\endhead
~ & \textbf{Training} & \textbf{Validate} & \textbf{Dice of Validate} &
\textbf{Test} & \textbf{Dice of Test} \\
Fold 0 & 575 & 144 & 0.9366 & 206 & 0.9138±0.1364 \\
Fold 1 & 575 & 144 & 0.9332 & 206 & 0.9102±0.1421 \\
Fold 2 & 575 & 144 & 0.9459 & 206 & 0.9083±0.1513 \\
Fold 3 & 576 & 143 & 0.9095 & 206 & 0.9065±0.1387 \\
Fold 4 & 575 & 144 & 0.9377 & 206 & 0.9038±0.1486 \\
\multicolumn{6}{@{}>{\raggedright\arraybackslash}p{(\columnwidth - 10\tabcolsep) * \real{1.0000} + 10\tabcolsep}@{}}{%
HighRes-Net Model: Optuna-selected hyperparameters} \\
~ & \textbf{Training} & \textbf{Validate} & \textbf{Dice of Validate} &
\textbf{Test} & \textbf{Dice of Test} \\
Fold 0 & 575 & 144 & 0.9275 & 206 & 0.8803±0.1845 \\
Fold 1 & 575 & 144 & 0.9301 & 206 & 0.9127±0.0808 \\
Fold 2 & 575 & 144 & 0.9374 & 206 & 0.8946±0.1635 \\
Fold 3 & 576 & 143 & 0.9142 & 206 & 0.8978±0.1140 \\
Fold 4 & 575 & 144 & 0.9323 & 206 & 0.8981±0.1480 \\
\bottomrule()
\end{longtable}

Based on these results, it can be inferred that the manually selected
hyperparameters for the HighRes-Net model yield better segmentation
performance than the Optuna-selected hyperparameters.

\hypertarget{results-4-comparison-of-the-different-scores-of-the-three-models-using-optuna.}{%
\section{Results 4: Comparison of the different scores of the three
models using
Optuna.}\label{results-4-comparison-of-the-different-scores-of-the-three-models-using-optuna.}}

By analyzing the provided table that compares the different scores of
the three models (UNet, VNet, and HighRes-Net) using Optuna, we can gain
insights into the performance of each model for fetal resting-state fMRI
brain segmentation. Table 6 presents the Dice scores for the validation
and test sets across five folds for each model. The results of the
analysis are as follows:

\begin{itemize}
\item
  UNet: The Dice scores ranged from 0.9181 to 0.9440 for the validation
  set, indicating good accuracy in segmenting the fetal brain. For the
  test set, the Dice scores ranged from 0.9060 to 0.9283, demonstrating
  consistent performance with competitive results.
\end{itemize}

VNet:The VNet model showed slightly higher Dice scores than the UNet
model.

For the validation set, the Dice scores ranged from 0.9228 to 0.9490,
indicating a higher level of accuracy in segmenting the fetal brain
compared to UNet.

For the test set, the dice scores ranged from 0.9079 to 0.9258,
indicating competitive performance.

\begin{itemize}
\item
  HighRes-Net: The HighRes-Net model demonstrated varied results, with
  slightly lower Dice scores compared to UNet and VNet. For the
  validation set, the dice scores ranged from 0.9142 to 0.9374, showing
  good accuracy but slightly lower than those of UNet and VNet. For the
  test set, the Dice scores ranged from 0.8588 to 0.9220, indicating a
  lower accuracy level than the other two models.
\end{itemize}

\begin{longtable}[]{@{}
  >{\raggedright\arraybackslash}p{(\columnwidth - 12\tabcolsep) * \real{0.1127}}
  >{\raggedright\arraybackslash}p{(\columnwidth - 12\tabcolsep) * \real{0.1479}}
  >{\raggedright\arraybackslash}p{(\columnwidth - 12\tabcolsep) * \real{0.1479}}
  >{\raggedright\arraybackslash}p{(\columnwidth - 12\tabcolsep) * \real{0.1479}}
  >{\raggedright\arraybackslash}p{(\columnwidth - 12\tabcolsep) * \real{0.1480}}
  >{\raggedright\arraybackslash}p{(\columnwidth - 12\tabcolsep) * \real{0.1479}}
  >{\raggedright\arraybackslash}p{(\columnwidth - 12\tabcolsep) * \real{0.1479}}@{}}
\caption{Table 6. Results of the HighRes-Net Model in fMRI Brain
Extraction}\tabularnewline
\toprule()
\begin{minipage}[b]{\linewidth}\raggedright
\end{minipage} &
\multicolumn{2}{>{\raggedright\arraybackslash}p{(\columnwidth - 12\tabcolsep) * \real{0.2957} + 2\tabcolsep}}{%
\begin{minipage}[b]{\linewidth}\raggedright
\textbf{UNet}
\end{minipage}} &
\multicolumn{2}{>{\raggedright\arraybackslash}p{(\columnwidth - 12\tabcolsep) * \real{0.2958} + 2\tabcolsep}}{%
\begin{minipage}[b]{\linewidth}\raggedright
\textbf{VNet}
\end{minipage}} &
\multicolumn{2}{>{\raggedright\arraybackslash}p{(\columnwidth - 12\tabcolsep) * \real{0.2957} + 2\tabcolsep}@{}}{%
\begin{minipage}[b]{\linewidth}\raggedright
\textbf{HighRes-Net}
\end{minipage}} \\
\midrule()
\endfirsthead
\toprule()
\begin{minipage}[b]{\linewidth}\raggedright
\end{minipage} &
\multicolumn{2}{>{\raggedright\arraybackslash}p{(\columnwidth - 12\tabcolsep) * \real{0.2957} + 2\tabcolsep}}{%
\begin{minipage}[b]{\linewidth}\raggedright
\textbf{UNet}
\end{minipage}} &
\multicolumn{2}{>{\raggedright\arraybackslash}p{(\columnwidth - 12\tabcolsep) * \real{0.2958} + 2\tabcolsep}}{%
\begin{minipage}[b]{\linewidth}\raggedright
\textbf{VNet}
\end{minipage}} &
\multicolumn{2}{>{\raggedright\arraybackslash}p{(\columnwidth - 12\tabcolsep) * \real{0.2957} + 2\tabcolsep}@{}}{%
\begin{minipage}[b]{\linewidth}\raggedright
\textbf{HighRes-Net}
\end{minipage}} \\
\midrule()
\endhead
& \textbf{manually} & \textbf{optuna} & \textbf{manually} &
\textbf{optuna} & \textbf{manually} & \textbf{optuna} \\
Fold 0 & 0.9260±0.0960 & 0.9282±0.0768 & 0.9215±0.1340 & 0.9277±0.0839 &
0.9138±0.1364 & 0.8803±0.1845 \\
Fold 1 & 0.9108±0.1296 & 0.9260±0.1140 & 0.9126±0.1522 & 0.9286±0.0812 &
0.9102±0.1421 & 0.9127±0.0808 \\
Fold 2 & 0.9025±0.1303 & 0.9228±0.1025 & 0.9131±0.1528 & 0.9189±0.1237 &
0.9083±0.1513 & 0.8946±0.1635 \\
Fold 3 & 0.9097±0.1034 & 0.9060±0.1018 & 0.9300±0.0829 & 0.9226±0.0886 &
0.9065±0.1387 & 0.8978±0.1140 \\
Fold 4 & 0.9035±0.1065 & 0.9219±0.1165 & 0.9355±0.0757 & 0.9181±0.1018 &
0.9038±0.1486 & 0.8981±0.1480 \\
\bottomrule()
\end{longtable}

Overall, the VNet model exhibited the highest Dice scores, followed by
the UNet and HighRes-Net models. These results suggest that both VNet
and UNet are effective models for fetal resting-state fMRI brain
segmentation, with VNet performing slightly better. HighRes-Net, despite
having slightly lower scores, still shows potential and should be
further explored for improvement.

\hypertarget{results-5-properties-of-best-hyperparameters-selected-by-optuna-for-each-model}{%
\section{Results 5: Properties of Best Hyperparameters selected by
Optuna for each
model}\label{results-5-properties-of-best-hyperparameters-selected-by-optuna-for-each-model}}

To further assess the performance of different models, we compared the
Properties in Hyperparameters selected by Optuna for each model (UNet,
VNet, and HighResNet). Table 3 provides insights into the training
duration, learning rate, dropout rate, and performance of each model.

\begin{itemize}
\item
  The models were trained on an NVIDIA GeForce RTX 2080 Ti graphics card
  with a memory capacity of 11264MiB.
\end{itemize}

\begin{longtable}[]{@{}
  >{\raggedright\arraybackslash}p{(\columnwidth - 14\tabcolsep) * \real{0.1533}}
  >{\raggedright\arraybackslash}p{(\columnwidth - 14\tabcolsep) * \real{0.1520}}
  >{\raggedright\arraybackslash}p{(\columnwidth - 14\tabcolsep) * \real{0.0948}}
  >{\raggedright\arraybackslash}p{(\columnwidth - 14\tabcolsep) * \real{0.1350}}
  >{\raggedright\arraybackslash}p{(\columnwidth - 14\tabcolsep) * \real{0.1246}}
  >{\raggedright\arraybackslash}p{(\columnwidth - 14\tabcolsep) * \real{0.0858}}
  >{\raggedright\arraybackslash}p{(\columnwidth - 14\tabcolsep) * \real{0.1234}}
  >{\raggedright\arraybackslash}p{(\columnwidth - 14\tabcolsep) * \real{0.1310}}@{}}
\caption{Table 7. Comparative Dice Scores for Fetal Resting State MRI
Brain Segmentation Models. This table shows the Dice scores obtained
from the five-fold validation and test sets for the UNet, VNet, and
HighRes-Net models, providing a comprehensive evaluation of their
performance.}\tabularnewline
\toprule()
\multirow{2}{*}{\begin{minipage}[b]{\linewidth}\raggedright
~
\end{minipage}} & \begin{minipage}[b]{\linewidth}\raggedright
\textbf{Time}
\end{minipage} &
\multirow{2}{*}{\begin{minipage}[b]{\linewidth}\raggedright
\textbf{epoch}
\end{minipage}} & \begin{minipage}[b]{\linewidth}\raggedright
\textbf{Learning}
\end{minipage} & \begin{minipage}[b]{\linewidth}\raggedright
\textbf{Weight}
\end{minipage} & \begin{minipage}[b]{\linewidth}\raggedright
\textbf{Step}
\end{minipage} & \begin{minipage}[b]{\linewidth}\raggedright
\textbf{Factor}
\end{minipage} & \begin{minipage}[b]{\linewidth}\raggedright
\textbf{Dropout}
\end{minipage} \\
& \begin{minipage}[b]{\linewidth}\raggedright
\textbf{consuming}
\end{minipage} & & \begin{minipage}[b]{\linewidth}\raggedright
\textbf{rate}
\end{minipage} & \begin{minipage}[b]{\linewidth}\raggedright
\textbf{Decay}
\end{minipage} & \begin{minipage}[b]{\linewidth}\raggedright
\textbf{Size}
\end{minipage} & \begin{minipage}[b]{\linewidth}\raggedright
\textbf{(gamma)}
\end{minipage} & \begin{minipage}[b]{\linewidth}\raggedright
\textbf{rate}
\end{minipage} \\
\midrule()
\endfirsthead
\toprule()
\multirow{2}{*}{\begin{minipage}[b]{\linewidth}\raggedright
~
\end{minipage}} & \begin{minipage}[b]{\linewidth}\raggedright
\textbf{Time}
\end{minipage} &
\multirow{2}{*}{\begin{minipage}[b]{\linewidth}\raggedright
\textbf{epoch}
\end{minipage}} & \begin{minipage}[b]{\linewidth}\raggedright
\textbf{Learning}
\end{minipage} & \begin{minipage}[b]{\linewidth}\raggedright
\textbf{Weight}
\end{minipage} & \begin{minipage}[b]{\linewidth}\raggedright
\textbf{Step}
\end{minipage} & \begin{minipage}[b]{\linewidth}\raggedright
\textbf{Factor}
\end{minipage} & \begin{minipage}[b]{\linewidth}\raggedright
\textbf{Dropout}
\end{minipage} \\
& \begin{minipage}[b]{\linewidth}\raggedright
\textbf{consuming}
\end{minipage} & & \begin{minipage}[b]{\linewidth}\raggedright
\textbf{rate}
\end{minipage} & \begin{minipage}[b]{\linewidth}\raggedright
\textbf{Decay}
\end{minipage} & \begin{minipage}[b]{\linewidth}\raggedright
\textbf{Size}
\end{minipage} & \begin{minipage}[b]{\linewidth}\raggedright
\textbf{(gamma)}
\end{minipage} & \begin{minipage}[b]{\linewidth}\raggedright
\textbf{rate}
\end{minipage} \\
\midrule()
\endhead
UNet & 4\textasciitilde5 hours & 100 & 9.40E-04 & 1.10E-05 & 8 & 0.7657
& 0.1554 \\
VNet & 5\textasciitilde6 hours & 60 & 3.40E-04 & 1.00E-05 & 20 & 0.7527
& 3.43E-04 \\
HighRes-Net & 25\textasciitilde26 hours & 60 & 1.00E-04 & 0 & 5 & 0.9 &
0 \\
\bottomrule()
\end{longtable}

\begin{itemize}
\item
  The UNet model was trained for 100 epochs with a learning rate of
  9.40E-04. It exhibited a dropout rate of 0.7657 and achieved promising
  results within a relatively short training time of 4--5 h.
\item
  Similarly, the VNet model underwent training for 60 epochs with a
  learning rate of 3.40E-04. This demonstrated a dropout rate of 0.7527
  and competitive performance. However, the training time was slightly
  longer, at approximately 5--6 h.
\item
  For our HighRes-Net architecture, a longer training time of 25 to 26
  hours was required for 60 epochs. Although no weight decay or dropout
  information was provided for this model, its architecture is worth
  considering owing to its potential benefits.
\end{itemize}

Overall, the UNet and VNet models exhibited efficient training times and
competitive performances, whereas the HighRes-Net model required a
longer training duration. These findings highlight the importance of
considering both training time and performance when selecting a suitable
model for a given task.

It is noteworthy that the results obtained from the Optuna-selected
hyperparameters are not explicitly mentioned in this table. However, it
is expected that the Optuna-selected hyperparameters would generally
yield better performance than the manually selected hyperparameters, as
demonstrated in the previous tables for the UNet and VNet models.

\hypertarget{results-6-hyperparameter-importance}{%
\section{\texorpdfstring{Results 6: Hyperparameter importance
}{Results 6: Hyperparameter importance }}\label{results-6-hyperparameter-importance}}

This section includes optimization history plots to provide a
comprehensive view of the hyperparameter optimization process for the
UNet, VNet, and HighRes-Net models. By analyzing these plots, readers
can gain a deeper understanding of the iterative refinement of
hyperparameters and the resulting impact on the performance of the
models.

\begin{itemize}
\item
  lr (Learning Rate): The rate at which the model learns during
  training.
\item
  Weight \_decay: A regularization technique that adds a penalty term to
  the loss function to prevent overfitting.
\item
  factors used in learning rate scheduling or decay methods.
\item
  dropout\_rate: Dropout is a regularization technique that randomly
  sets a fraction of input units to 0 at each update during training.
\item
  Step \_size: Step size used in learning rate scheduling methods.
\item
  Dice: The Dice coefficient is a common evaluation metric for image
  segmentation tasks, which measures the similarity between the
  predicted and ground truth masks.
\end{itemize}

Here are the hyperparameters that resulted in the best dice score for
each of the three models, based on Optuna Selection:

UNet:

\begin{itemize}
\item
  Learning Rate (lr): 0.000602891
\item
  Weight Decay: 2.22E-05
\item
  Factor: 0.812967388
\item
  Dropout Rate: 0.042077713
\item
  Step Size: 12
\item
  Dice Score: 0.940994442
\end{itemize}

V-Net:

\begin{itemize}
\item
  Learning Rate (lr): 0.000338659
\item
  Factor: 0.752687201
\item
  Dropout Rate: 0.00034294
\item
  Step Size: 20
\item
  Dice Score: 0.950378239
\item
  Weight Decay: 1.00E-05
\end{itemize}

HighRes-Net:

\begin{itemize}
\item
  Learning Rate (lr): 0.000414447
\item
  Weight Decay: 1.53E-05
\item
  Dropout Rate: 0.116149873
\item
  Dice Score: 0.928082108
\item
  Factor: 0.19944579
\item
  Step Size: 16
\item
\end{itemize}

\hypertarget{result-6.1-optimization-history-plot}{%
\section{\texorpdfstring{Result 6.1: Optimization History Plot
}{Result 6.1: Optimization History Plot }}\label{result-6.1-optimization-history-plot}}

The optimization history plot provides valuable insights into the
gradual optimization process of hyperparameters for the different models
(UNet, VNet, and HighRes-Net).

The optimization history plot shows the progression of the
hyperparameter search over iterations or epochs. It visually represents
the evolution of the hyperparameters and the corresponding performance
metrics achieved during optimization.

In the figures below, for each model (Unet, VNet, and HighRes-Net), the
optimization history plot shows the changes in hyperparameters during
the trials. Trianl numbers were considered for UNet:100, VNet:20, and
HighRes-Net:10. The reason for this difference is that the HighRes-Net
and VNet models were too time-consuming compared to the UNet model. The
plots also display the learning rate, weight decay, and dropout rate
over iterations.

As shown in the figure below, the Dice score increased in the UNet
trials from 0 to 100. In addition, the Dice net score showed an
increasing trend in the VNet model. However, HighRes-Net did not exhibit
the same trend in the ten trials. We hypothesized that if we had more
trials for the HighRess-Net model, an increasing trend would be seen (we
did not increase the trial numbers of HighRes-Net to make the time
consumed equal between the different models. We did that to compare in
fair situation for clinical use)

\begin{itemize}
\item
  UNet
\end{itemize}

\includegraphics[width=6.5in,height=1.83019in]{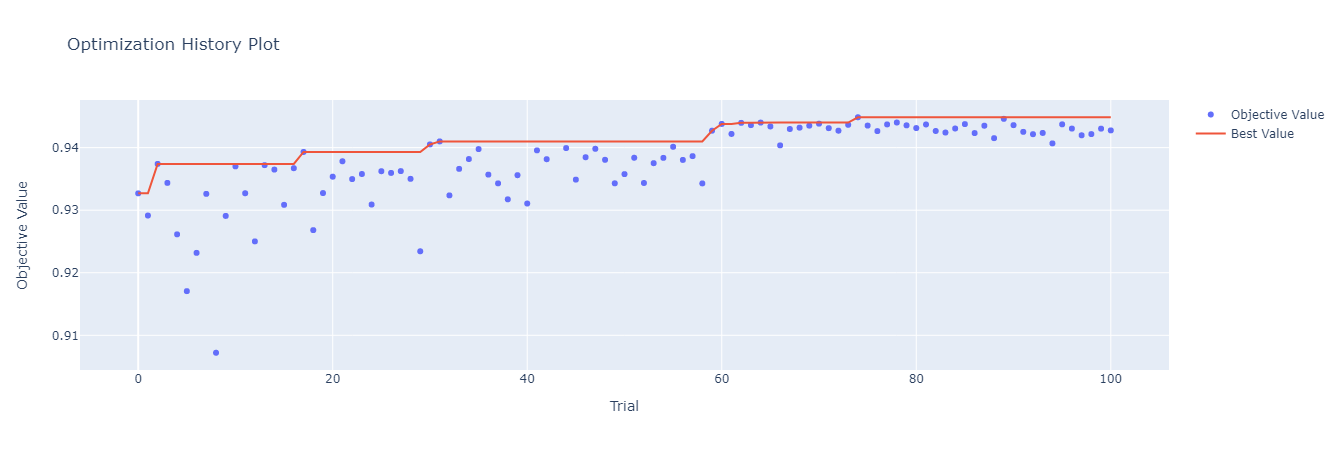}

\begin{itemize}
\item
  VNet
\end{itemize}

\includegraphics[width=6.5in,height=1.58681in]{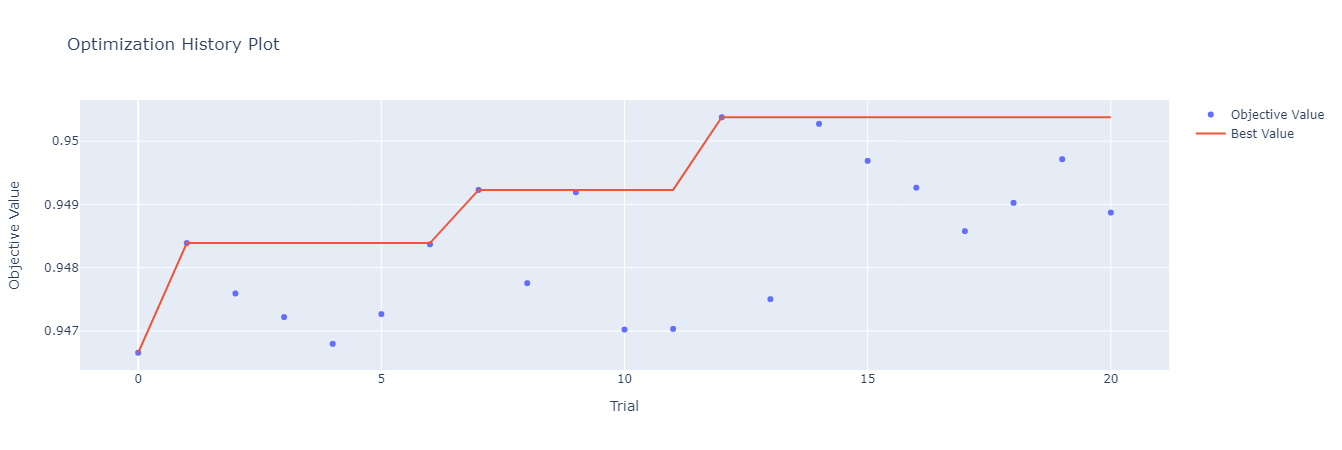}

\begin{itemize}
\item
  HighRes-Net
\end{itemize}

\includegraphics[width=6.5in,height=1.65903in]{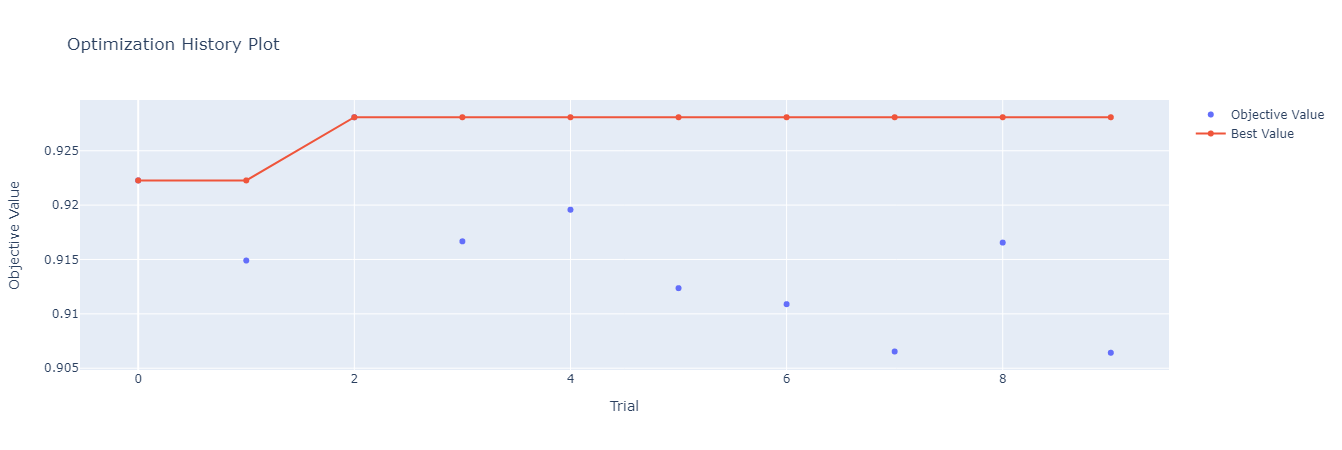}

Figure 5- Optimization History Plot for three AI models for fMRI fetal
segmentation

Figure: optimization history plot for three different models

These plots are crucial for understanding the optimization journey and
how the model performance improves as the hyperparameters are adjusted.
They provide valuable insights into the effectiveness of the
optimization process and help identify the optimal set of
hyperparameters for each model.

\hypertarget{results-6.2-hyperparameter-importance}{%
\section{Results 6.2: Hyperparameter
Importance}\label{results-6.2-hyperparameter-importance}}

This section discusses the importance of the hyperparameters by
identifying the most influential hyperparameters for each model.
Analyzing the importance of hyperparameters is crucial for understanding
the impact of different settings on the performance of models, including
UNet, VNet, and HighRes-Net.

Figure 6 shows that we used this function from the optuna.

\emph{fig = optuna.visualization.plot\_param\_importances(study)}

In general, the importance of hyperparameters can be determined through
various methods, such as sensitivity analysis, feature importance
ranking, and analysis of their effects on the model\textquotesingle s
performance. These techniques help identify the hyperparameters that
have the most significant influence on the model\textquotesingle s
accuracy, convergence, and generalization capabilities.

For the UNet, VNet, and HighRes-Net models, analyzing hyperparameter
importance involves assessing the impact of key parameters, such as
learning rate, weight decay, dropout rate, and architecture-specific
parameters, on the model\textquotesingle s performance. By
systematically varying these hyperparameters and measuring their effects
on validation and test scores, researchers can gain insights into the
relative importance of each parameter to achieve optimal results.

For each model, the most important factors were:

\begin{itemize}
\item
  For Unet: "lr" and "factor" factors.
\item
  For Vnet: "Factor" and "lr"
\item
  For UNet: ``droupout\_rate'' and ``weight\_decay''
\end{itemize}

Unet

\includegraphics[width=6.5in,height=1.58889in]{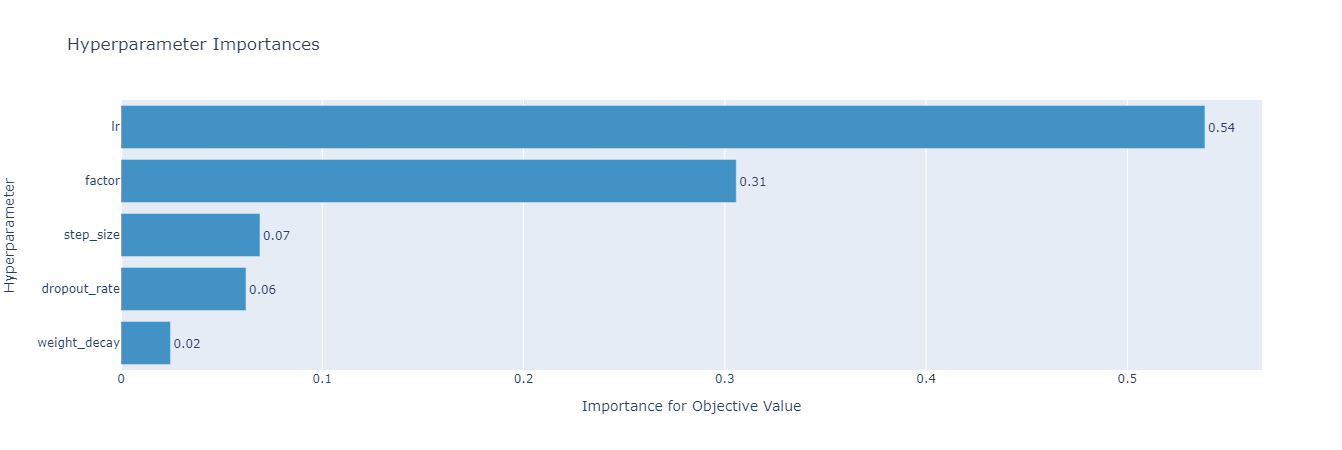}

VNet

\includegraphics[width=6.5in,height=1.58889in]{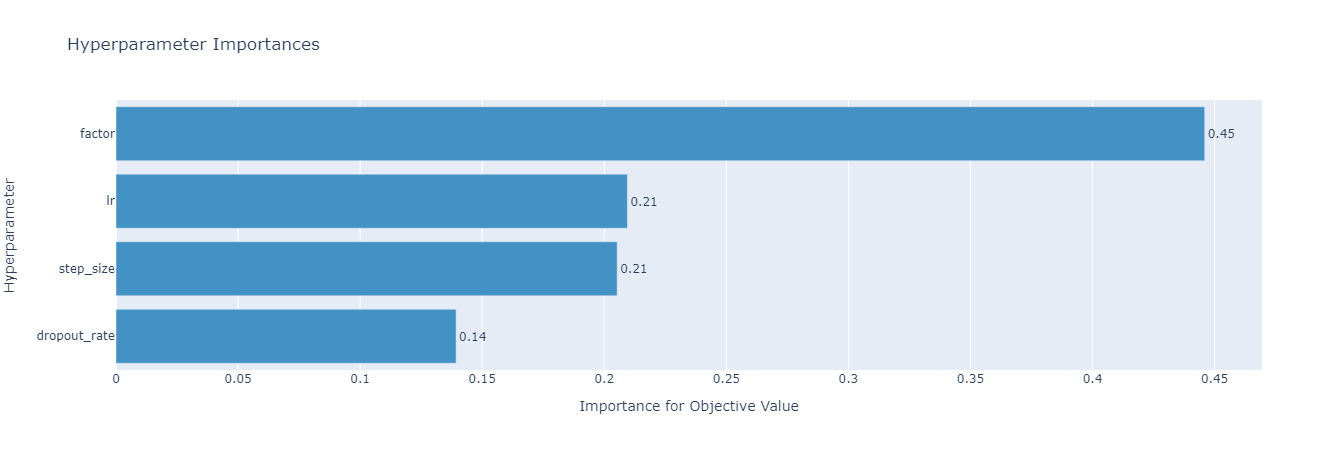}

HigRes-Net

\includegraphics[width=6.5in,height=1.58889in]{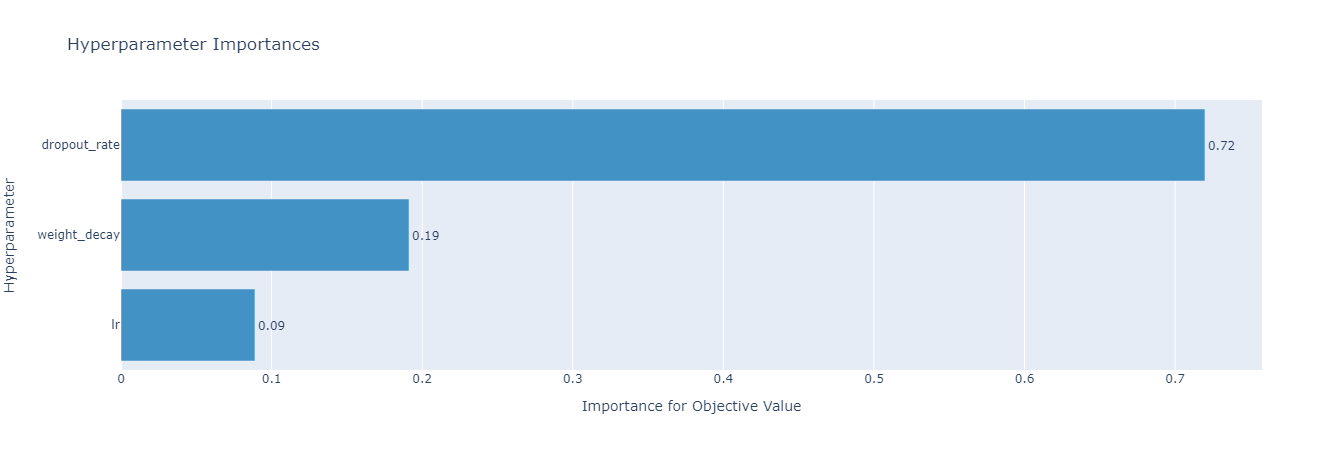}

Figure 6- Analysis of Hyperparameter Importance for three AI models for
fMRI fetal segmentation

\hypertarget{results-6.3-slice-plot-of-different-parameters}{%
\section{Results 6.3: Slice Plot of different
parameters}\label{results-6.3-slice-plot-of-different-parameters}}

This section shows the influence of different parameters (dropout\_rate,
lr, etc.) on the optimization results.

The slice plot is a visual representation that illustrates the impact of
different parameters on the optimization results. It typically shows how
changing individual hyperparameters, such as dropout and learning rates,
affects the performance metrics of models.

In the context of the UNet, VNet, and HighRes-Net models, the slice plot
demonstrates the influence of various parameters on the optimization
results for each model. By plotting different parameter values along one
axis and the corresponding performance metrics (e.g., accuracy, loss, or
validation score) along the other axis, the slice plot provides a visual
examination of how parameter variations affect the performance of the
models.

The figure below shows the concentration of each factor in the Dice
score. For example, in the UNet model, the concentration of
"dropout\_rate" with a high dice score was between 0.1 and 0.2.
Similarly, the highest Dice score in the UNet model belonged to the "lr"
between 0.7 and 0.9.

UNet

\includegraphics[width=6.5in,height=1.58889in]{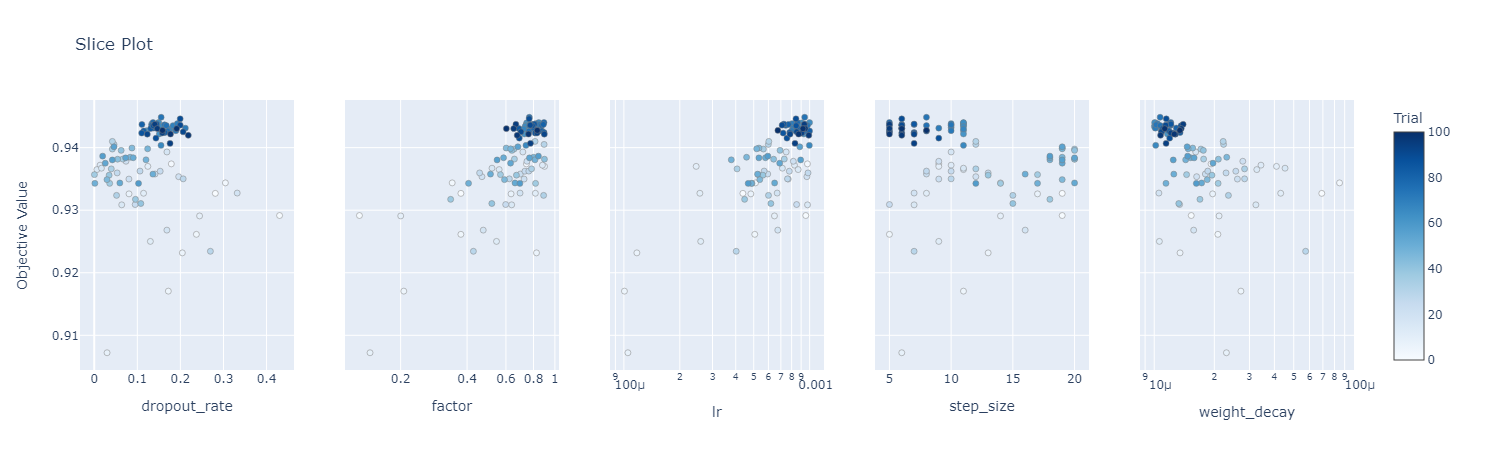}

VNet

\includegraphics[width=6.5in,height=1.58889in]{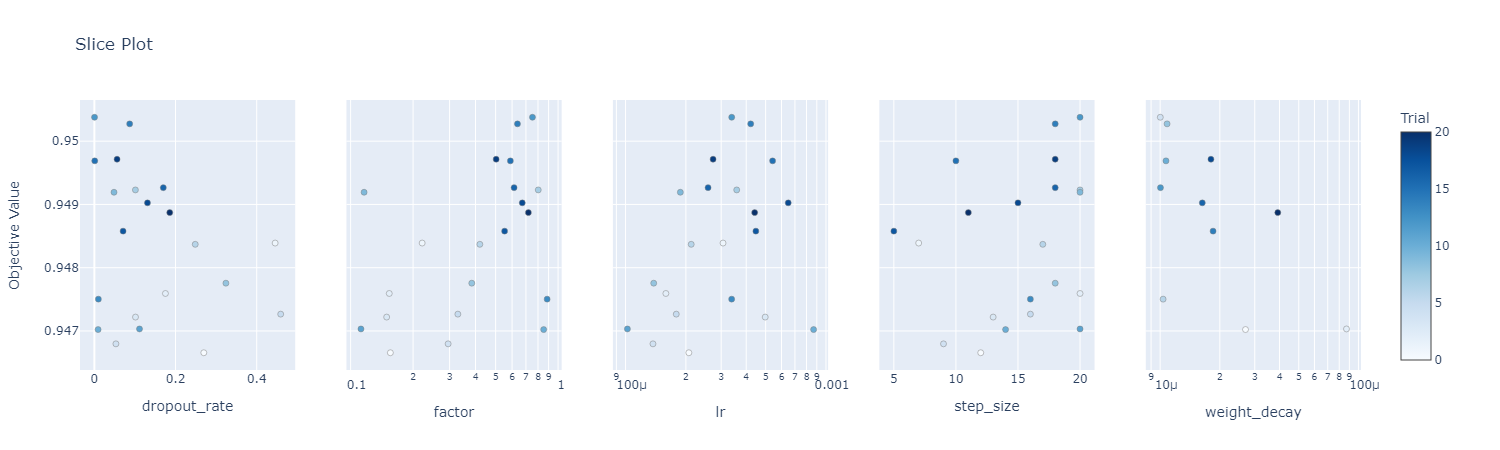}

HighRes-Net

\includegraphics[width=6.5in,height=1.58889in]{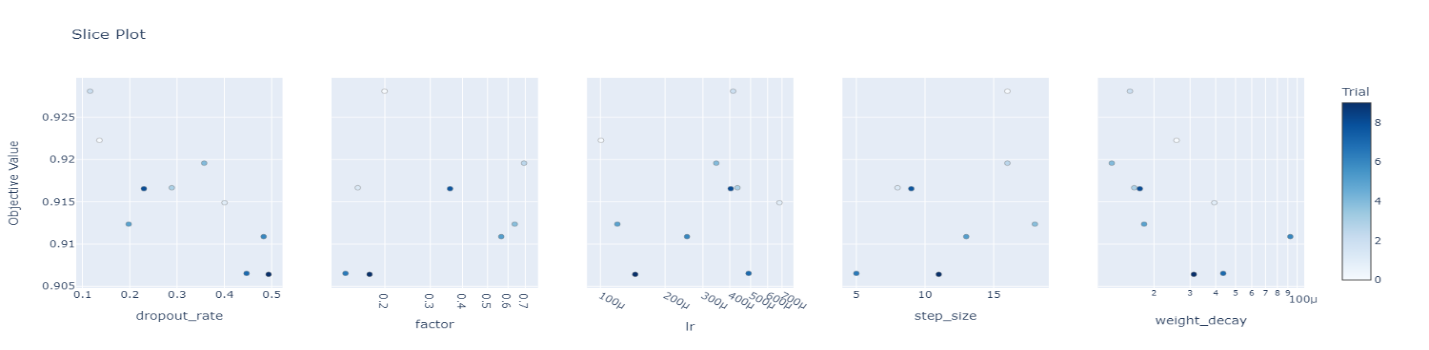}

Figure 7- Slice Plot of different parameters for three AI models for
fMRI fetal segmentation

\hypertarget{results-6.4-contour-plot-of-different-parameters}{%
\section{Results 6.4: Contour Plot of different
parameters}\label{results-6.4-contour-plot-of-different-parameters}}

Contour plots are presented in this section. By examining the contour
plots, insight can be gained into the optimal ranges or values for these
hyperparameters and how they impact the convergence, generalization, or
other relevant performance measures of the models.

UNet

\includegraphics[width=6.5in,height=1.58819in]{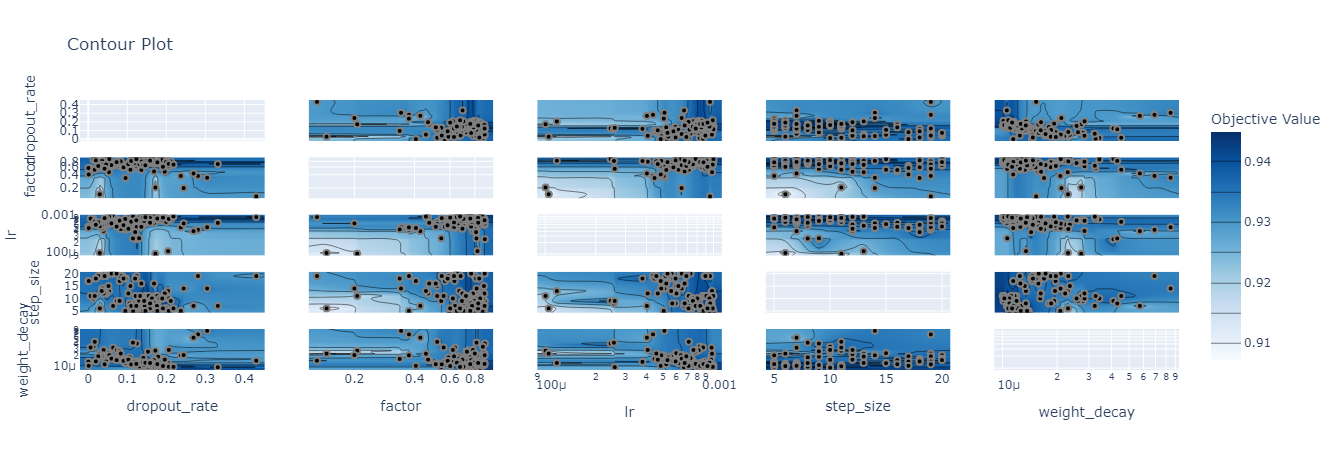}

VNet

\includegraphics[width=6.67928in,height=1.632in]{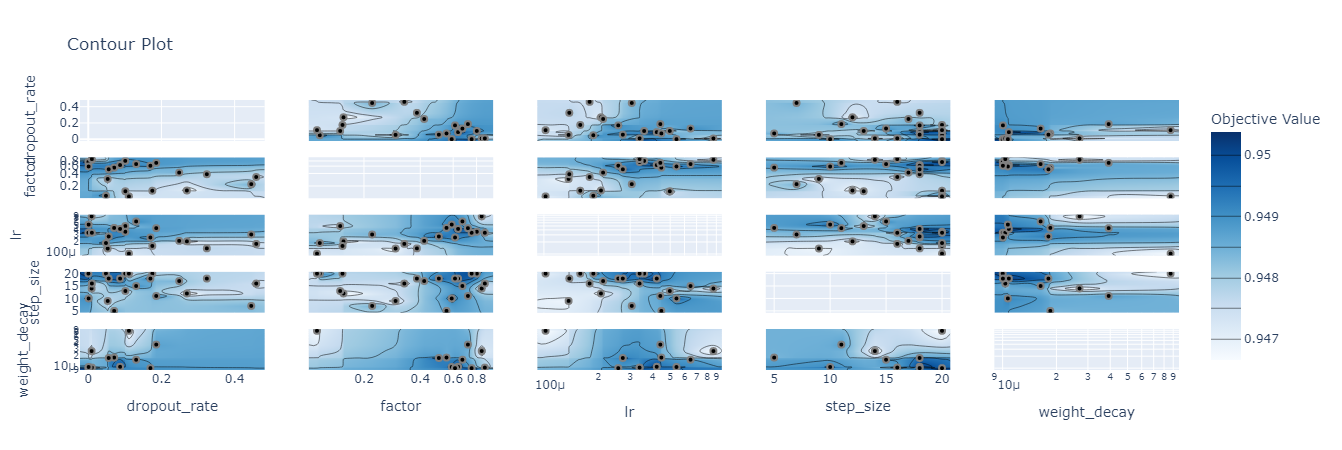}

HighRes-Net

\includegraphics[width=6.5in,height=1.56458in]{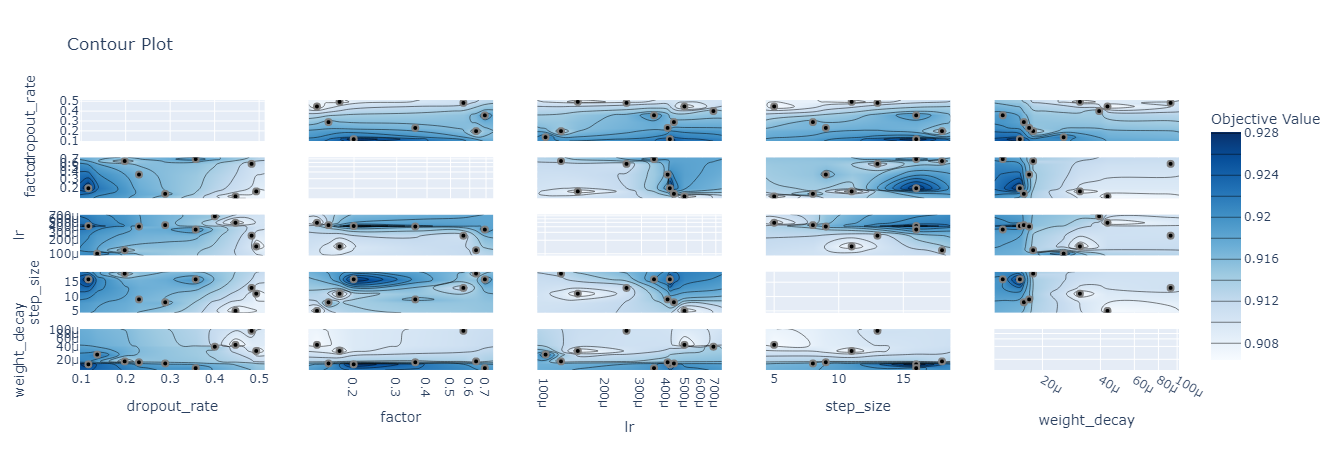}

Figure 8- Contour plots for three AI models for fMRI fetal segmentation

\textbf{Other Preprocessing}

After the fetal brain has been extracted, the data can enter more
typical preprocessing steps, for which child/adult tools have been
developed. However, the problem of increased head motion in fetal data
remains a challenge. The correct threshold for censoring high motion
volumes from a BOLD time series is an ongoing topic of discussion. We
tested various censoring thresholds, and the resulting amounts of data
preserved at each censoring threshold are shown in Figure.

In conclusion, our AI models demonstrated promising performance in
automating fMRI segmentation of the fetal brain. However, further
research is needed to improve image quality and overcome head motion
challenges to leverage the full potential of these models.

\textbf{Age \& Data Quality Failure Analysis}

We observed a significant positive correlation between gestational age
and the Dice coefficient, suggesting that our models performed better on
older fetuses. However, we noted that the performance of our models
decreased when the brain was significantly displaced from the image
origin or when the image quality was poor.

\hypertarget{discussion}{%
\section{Discussion:}\label{discussion}}

We utilized an open-source fetal functional MRI (fMRI) dataset
consisting of 160 cases (reference: fetal-fMRI - OpenNeuro). An AI model
for fMRI segmentation was developed using a 5-fold cross-validation
methodology. Three AI models were employed: 3D UNet, VNet, and
HighResNet. Optuna, an automated hyperparameter-tuning tool, was used to
optimize these models.

We compared the Dice scores from the three AI models (Vnet, UNet, and
HighRes-Net). In addition, we compared the model with manual and
automatic tuning (Optuna).

Regarding the architecture of the models, all three showed high accuracy
in medical image segmentation tasks, including the segmentation of fetal
brain structures in rs-fMRI. They differ in their architectural designs,
training strategies, and regularization techniques. The UNet and VNet
models utilize encoder-decoder architectures with different types of
connections (skip connections and residual connections, respectively),
whereas the HighRes-Net model employs a 3D convolutional network. The
models used various evaluation metrics and training strategies,
requiring GPUs with sufficient memory for computational resources. In
addition, dropout, weight decay, and batch normalization were used as
regularization techniques for different combinations across the models.

It is worth noting that the performance of these models can vary across
different folds, which may indicate some level of variability in the
data or the sensitivity of the model to different training instances.
However, a more comprehensive analysis would require additional
statistical measures or comparisons to draw definitive conclusions
regarding the superiority of one model over the others.

These results provide insights into the performance of different AI
models for fetal resting-state fMRI brain segmentation and can serve as
a basis for further research and development in this area.

The performance analysis of the UNet model demonstrated competitive
results for both the manually selected and Optuna-selected
hyperparameters. The model achieved high Dice scores, ranging from
0.9114 to 0.9440 for the validation set and from 0.9025 to 0.9282 for
the test set. The Optuna-selected hyperparameters further improved the
model\textquotesingle s performance, leading to slightly higher Dice
scores across the validation and test sets.

Similarly, the VNet model exhibited strong segmentation performance for
manually selected and Optuna-selected hyperparameters. The Dice scores
ranged from 0.9254 to 0.9521 for the validation set and from 0.9126 to
0.9355 for the test set. Once again, the Optuna-selected hyperparameters
slightly improved the dice scores compared with the manually selected
ones.

In contrast, the HighRes-Net model demonstrated varied performances for
both manually selected and Optuna-selected hyperparameters. The Dice
scores ranged from 0.9095 to 0.9459 for the validation set and 0.8803 to
0.9138 for the test set. Notably, the Optuna-selected hyperparameters
did not yield higher Dice scores compared to the manually selected
hyperparameters.

A comparison of different scores using Optuna across the three models
reaffirmed the competitive performance of UNet and VNet. The VNet model
consistently achieved higher Dice scores than UNet, indicating its
potential superiority in fetal rs-fMRI brain segmentation. The
HighRes-Net model, while exhibiting slightly lower Dice scores, still
showed potential and could be further explored for improvement.

The application of rs-fMRI to the fetal brain presents numerous
challenges that are currently impeding progress in this field. These
challenges are primarily due to the unique characteristics of the fetal
brain and surrounding environment. Unlike the adult brain, which is
relatively stationary and isolated from other bodily tissues, the fetal
brain is constantly moving and encased within the maternal body. This
leads to significant signal noise and variability that traditional fMRI
analysis tools designed with adult and child data in mind are not
equipped to handle.

Additionally, accurate segmentation of the fetal brain from the
surrounding tissues is a critical challenge in fetal fMRI analysis. The
process of distinguishing brain tissue from non-brain tissue is
complicated by the movement of the fetus, the small size of the fetal
brain, and the presence of maternal tissue around the fetus. Currently,
the most commonly used approach to achieve acceptable standards is the
manual generation of brain masks, which is tedious, time-consuming, and
prone to human errors and inconsistencies.

The development of AI models for fetal resting-state functional magnetic
resonance imaging (rs-fMRI) brain segmentation represents a significant
milestone in neuroscience and developmental biology. While traditional
MRI analysis tools have been predominantly designed for adult and child
data, the unique challenges posed by fetal brain imaging necessitate the
development of more specialized tools. This study, which introduces a
novel application of AI for automated brain segmentation in fetal brain
fMRI, constitutes a pioneering step in this direction.

The application of artificial intelligence to brain imaging and
segmentation has gained traction in recent years. However, most of these
studies have focused on adult or postnatal brain images. Dolz et al.
(2018) developed a 3D fully convolutional network to segment brain
lesions using multimodal MRI scans. (15) Similarly, Yamanakkanavar et
al. (2018) proposed a deep learning model for segmenting brain MRI
scans, with a focus on identifying brain lesions. (16) While these
studies have demonstrated the efficacy of AI in brain imaging, applying
these models to fetal brain fMRI presents a new set of challenges unique
to this domain.

A comparative analysis of the three AI models utilized in the present
study -- the 3D UNet, VNet, and HighRes-Net models -- revealed
interesting insights into their strengths and limitations. The VNet
model, initially introduced by Milletari et al. (14) (17) for volumetric
medical image segmentation, outperformed both the 3D UNet and
HighRes-Net models in fetal brain fMRI segmentation. The superior
accuracy of the VNet model can be attributed to its ability to capture
and process high-resolution features from the input images, which is
crucial for accurately segmenting complex structures, such as the fetal
brain.

However, the 3D UNet model, an adaptation of the original UNet model
proposed by Ronneberger et al. (2015), showed consistent performance
despite its slightly lower accuracy compared to the VNet model. (18) The
3D UNet model is known for its robustness and efficiency in biomedical
image segmentation, and our findings confirmed its potential
applicability in the field of fetal brain fMRI.

The HighRes-Net model, despite its lower performance in this study, may
have potential advantages in other aspects not covered in this study.
For instance, Li et al. applied the HighRes-Net model to a brain MRI
white matter hyperintensity segmentation task and achieved promising
results (14)(17). Although the HighRes-Net model did not perform as well
as the VNet and 3D UNet models in our study, it is plausible that
certain modifications or adaptations could enhance its performance in
fetal brain fMRI segmentation.

In the context of similar studies, our research presents a novel
approach for tackling the unique challenges of fetal brain fMRI
segmentation. Gholipour et al. (2017) developed a convolutional neural
network for fetal brain MRI segmentation focusing primarily on
structural MRI data. (19) Our study extends this line of research by
applying AI models to functional MRI data, which introduces additional
complexities owing to the dynamic nature of the fetal brain.

Salehi et al. (2020) compared several machine learning models for brain
MRI segmentation, including the UNet model. (20) However, this study
focused on images of the adult brain and used 2D models. Our research
contributes to this field by employing 3D models that are more suitable
for handling the volumetric nature of fMRI data.

Current tools also have low accuracy for fetal rain extraction in fMRI
in the current dataset. BET (Brain Extraction tool) from FSL has dice
score of 0.22 (+/- 0.13), and 3dSS ( 3dSkullStrip tool ) from AFNI had
0.24 (+/- 0.10). (21)

\hypertarget{conclusion}{%
\section{Conclusion:}\label{conclusion}}

This study demonstrated the potential of AI models for fetal rs-fMRI
brain segmentation, particularly the VNet model. The findings highlight
the importance of considering both the training time and performance
when selecting a suitable model. The use of Optuna for hyperparameter
optimization resulted in improvements in model performance and
emphasized the effectiveness of automated tuning methods. Further
research and development are encouraged to address the challenges posed
by fetal brain imaging and to explore enhancements of the HighResNet
model. Overall, this study provides valuable insights and directions for
future research in the field of fetal rs-fMRI.

These findings indicate that the VNet model shows promising results for
this application. Further research is needed to fully explore

accuracy and speed. The VNet model particularly stood out in terms of
performance, although all the models showed commendable results.

It is interesting to note the correlation between gestational age and
Dice coefficient. This finding could potentially inform future research
and development in this area. This suggests that the models might be
more reliable in processing images of older fetuses. This could be
because the brains of older fetuses are more developed, which provides
more distinct features for AI models to process and analyze.

The challenges posed by image quality and head motion are significant
and remain an area for further research. It would be interesting to
explore the potential of integrating advanced motion-correction
algorithms into AI models to enhance their performance in the face of
these challenges.

In terms of computation, it is noteworthy that the training and testing
times of the models were efficient, particularly when a GPU was used.
This suggests that these models are accurate and practical for
large-scale application.

Finally, exploring various censoring thresholds in the preprocessing
stage indicated that this area would benefit from further research.
Determining the correct censoring threshold could potentially enhance
data quality and AI model performance.\\
As detailed in our resources and tutorial, the proposed pipeline
effectively addresses the challenges associated with fetal brain fMRI
segmentation, and significantly reduces the manual effort required. We
hope that this pipeline will be instrumental in advancing the field of
fetal brain fMRI analysis, opening new avenues for understanding brain
development before birth.

Overall, this study presents an encouraging picture of the potential of
AI in fMRI segmentation of the fetal brain. The findings provide a solid
foundation for future research and a clear direction for further
development.

\hypertarget{references}{%
\section{References:}\label{references}}

1. Moore R, Vadeyar S, Fulford J, Tyler D, Gribben C, Baker P, et al.
Antenatal determination of fetal brain activity in response to an
acoustic stimulus using functional magnetic resonance imaging. Hum Brain
Mapp. 2001;12(2):6.

2. J F, SH V, SH D, RJ M, P Y, PN B, et al. Fetal brain activity in
response to a visual stimulus. Human brain mapping. 2003;20(4).

3. KK K, JW B, DA C, IE G, RM W, BP P, et al. Dynamic magnetic resonance
imaging of human brain activity during primary sensory stimulation.
Proceedings of the National Academy of Sciences of the United States of
America. 1992;89(12).

4. S O, DW T, R M, JM E, SG K, H M, et al. Intrinsic signal changes
accompanying sensory stimulation: functional brain mapping with magnetic
resonance imaging. Proceedings of the National Academy of Sciences of
the United States of America. 1992;89(13).

5. C M, KR D, PA G, IR J, PN B. Failure to detect intrauterine growth
restriction following in utero exposure to MRI. The British journal of
radiology. 1998;71(845).

6. A K, J H, E S, H P, S M, JM P, et al. Activation of multiple cortical
areas in response to somatosensory stimulation: combined
magnetoencephalographic and functional magnetic resonance imaging. Human
brain mapping. 1999;8(1).

7. Sharon D, Hamalainen MS, Tootell RB, Halgren E, Belliveau JW. The
advantage of combining MEG and EEG: comparison to fMRI in
focally-stimulated visual cortex. Neuroimage. 2007;36(4):11.

8. PI T, M K, V J, JP U, R H, R S, et al. Comparison of BOLD fMRI and
MEG characteristics to vibrotactile stimulation. NeuroImage. 2003;19(4).

9. Vahedifard F, Adepoju JO, Supanich M, Ai HA, Liu X, Kocak M, et al.
Review of deep learning and artificial intelligence models in fetal
brain magnetic resonance imaging. World. 2023;11(16):0-.

10. Vahedifard F, Ai HA, Supanich MP, Marathu KK, Liu X, Kocak M, et al.
Automatic Ventriculomegaly Detection in Fetal Brain MRI: A Step-by-Step
Deep Learning Model for Novel 2D-3D Linear Measurements. Diagnostics.
2023;13(14):2355.

11. Vahedifard F, Liu X, Marathu KK, Kocak M, Ai HA, Supanich MP, et al.
Artificial Intelligence Prediction of Gestational Age of Fetal in Brain
Magnetic Resonance Imaging versus ultrasound Using three different
Biometric Measurements. 2023.

12. Farzan Vahedifard JD, Mark Supanich, Jubril Adepoju, Xuchu Liu,
Sharon Byrd MD, editor Deep Learning Model for Automatic Landmark
Localization in Fetal Brain MRI. Annual Medical Education Conference
(AMEC), Florida, USA; 2022.

13. Kerfoot E, Clough J, Oksuz I, Lee J, King AP, Schnabel JA.
Left-Ventricle Quantification Using Residual U-Net \textbar{}
SpringerLink. 2023.

14. Li W, Wang G, Fidon L, Ourselin S, Cardoso MJ, Vercauteren T. On the
Compactness, Efficiency, and Representation of 3D Convolutional
Networks: Brain Parcellation as a Pretext Task. 2017.

15. J D, C D, I BA. 3D fully convolutional networks for subcortical
segmentation in MRI: A large-scale study. NeuroImage. 2018;170.

16. Yamanakkanavar N, Choi JY, Lee B. MRI Segmentation and
Classification of Human Brain Using Deep Learning for Diagnosis of
Alzheimer's Disease: A Survey. Sensors (Basel). 2020;20(11):1.

17. Milletari F, Navab N, Ahmadi S-A. V-Net: Fully Convolutional Neural
Networks for Volumetric Medical Image Segmentation. 2016.

18. Ronneberger O, Fischer P, Brox T. U-Net: Convolutional Networks for
Biomedical Image Segmentation. 2015.

19. Gholipour A, Rollins CK, Velasco-Annis C, Ouaalam A, Akhondi-Asl A,
Afacan O, et al. A normative spatiotemporal MRI atlas of the fetal brain
for automatic segmentation and analysis of early brain growth.
Scientific Reports. 2017;7(1):1-13.

20. A S, SSM S, A G. Deep Predictive Motion Tracking in Magnetic
Resonance Imaging: Application to Fetal Imaging. IEEE transactions on
medical imaging. 2020;39(11).

21. S R, P S, M A, J H, J W, MI vdH, et al. Automated Brain Masking of
Fetal Functional MRI with Open Data. Neuroinformatics. 2022;20(1).

\end{document}